\pdfoutput=1

\documentclass[
 reprint,       
 amsmath,
 amssymb,
 aps,           
 nofootinbib    
]{revtex4-2}

\usepackage{graphicx} 
\usepackage{dcolumn}  
\usepackage{bm}       
\usepackage{hyperref} 
\hypersetup{
    colorlinks=true,
    linkcolor=blue,
    filecolor=magenta,      
    urlcolor=blue,
    citecolor=blue,
}

\renewcommand{\d}{\text{d}}

\begin{document}

\title{Frustrated shapes of solid domains in fluid vesicle membranes:  From rolls and folds to crumples and wrinkles}

\author{Geunwoong Jeon}
\affiliation{Department of Physics, University of Massachusetts Amherst, United States}
\author{Anthony N. A. Prempeh}
\affiliation{Department of Chemical and Biomolecular Engineering, University of Massachusetts Amherst, United States}
\author{Maria M. Santore}
\author{Gregory M. Grason}
\email{grason@umass.edu}
\affiliation{Department of Polymer Science and Engineering, University of Massachusetts Amherst, United States}

\date{\today}

\begin{abstract}
Fluid-solid composite vesicles, comprising 2D solid domains integrated into a topologically closed otherwise fluid lipid bilayer membrane, exhibit complex morphologies arising from the geometric frustration between spherical closure of the membrane and 2D solid elasticity. This scenario is distinct from the better studied case of multi-fluid domain vesicles.  Here, we study the elastic energies and shape equilibria of a closed vesicle membrane containing a single, flexible circular solid domain using discrete finite-element (Surface Evolver) simulations, determining the key physical and mechanical parameters to govern shape selection.  While we find that the 2D solid (shear) elasticity has minimal impact on the highly-under inflated morphologies, the geometrically non-linear resistance of the 2D solid to Gaussian curvature substantially impacts the shape and elastic patterns form for inflated vesicles, by an amount that it grows with ratio of vesicle size to the elastic thickness of solid.  For sufficiently large (thin) vesicles we characterize a generic sequence of ground state patterns of solid shape with increasing inflation: from cylindrical rolls and isometric folds to spatially complex patterns of crumples and wrinkles and ultimately to smooth caps.  This sequence of non-isometric patterns at high-inflation is shown to be governed by the same far-from-threshold mechanics used to describe similar shape transitions in microscopic sheets on curved liquid interfaces, establishing that inflated shapes are governed by two basic mechanical scales of membrane tension.  We find our predictions for highly-anisotropic shape equilibria of fluid-solid composite vesicles closely match experimentally observed shapes of giant unilamellar vesicles of phase-separated DPPC and DOPC.  Our study reveals new connection between multi-component vesicles and the geometrically non-linear mechanics of thin solids, and more broadly suggests new principles for pattern control and selection in this paradigmatic class of biological soft matter.
\end{abstract}

\maketitle


\section{Introduction}

Thin 2D solids, from macroscopic paper, polymer films and growing leaves to 2D molecular sheets like graphene~\cite{blair2005geometry,kramer1996universal,lobkovsky1997properties,lv2014origami,payamyar2016two,payamyar2016two,liang2009shape,zhang2025geometrically,armon2011geometry,deng2016wrinkled,wang2013graphene,yllanes2017thermal}, are subject to generic and strongly non-linear interplay mechanics and geometry.  For 2D solids that are much thinner than their lateral dimensions, elasticity generically implies they are far more easily bent than stretched or deformed in-plane~\cite{witten2007stress}.  This underlies a host of elastic instabilities of 2D sheets.  Under different states external forcing or confinement 2D solids exhibit a rich spectrum of patterns, often characterized by fine structural features -- like wrinkles, crumples, folds or ridges~\cite{cerda2003geometry,witten2007stress,king2012elastic,paulsen2016curvature} -- whose spatial arrangement is non-linearly sensitive to loading.  One prototypical scenario that drives complex elastic patterning might be called \textit{geometrically incompatible confinement}~\cite{davidovitch2019geometrically,box2023delamination}, when a 2D solid forced to conform to surface shapes whose Gaussian curvatures conflict with 2D metric geometry of the solid.  This has been studied, for example, for flat sheets wrapping liquid drops or stamped between two spherical lenses, as well as for thin spherical or hyperbolic shells floating on flat liquid interfaces~\cite{paulsen2015optimal,hure2012stamping,tobasco2022exact}.  These diverse examples reveal generic principles underlying the organization of the complex elastic patterning of solids.  For example, for sufficiently thin sheets, wrinkles emerge to collapse and eliminate unstable compressive stress in the solid, and accordingly, their structure (spatial location, wavelength and amplitude) follow predictable, well-defined conditions~\cite{witten2007stress,davidovitch2011prototypical,king2012elastic}.  In this Article, we study the elastic patterns that result from a variant of this geometric frustration of 2D solids in paradigmatic class of soft and biological materials, lipid bilayer vesicles.

\begin{figure*}[t!]
\begin{center}
    \includegraphics[width=0.75\textwidth]{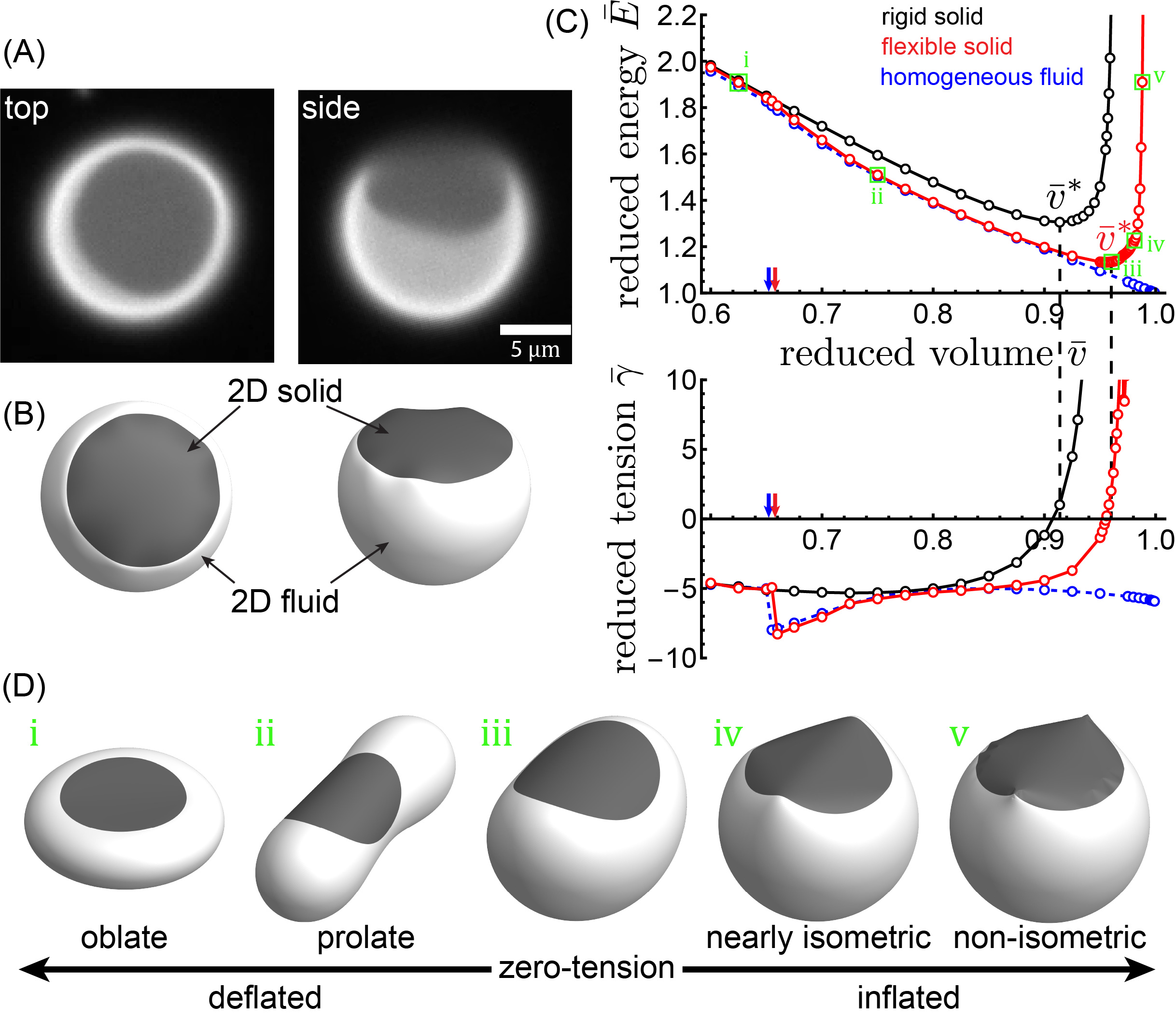}
\caption{\label{fig: figure1}\textbf{Complex equilibria of fluid-solid composite vesicles.} (A) A fluorescence micrograph of a fluid-solid (DOPC-DPPC) composite vesicle, following methods of ref.~\cite{wan2026sculpting}. (B) Surface Evolver model of a vesicle with a circular solid domain. (C) Plots of reduced energy and reduced tension (defined in text) as functions of reduced volume $\bar{v}$ for homogeneous fluid vesicles~\cite{seifert1991shape}, vesicles with a rigid solid domain~\cite{jeon2024shape}, and vesicles with a flexible solid domain, both with solid fraction $\Phi=0.15$. Blue and red arrows denote the transition point of the vesicle shape branch from oblate to prolate for homogeneous fluid vesicles and vesicles with flexible domains, respectively. (D) Example sequential structures for different regimes across the reduced volume range, labeled as (i) - (v) in (C).}
\end{center}
\end{figure*}

Lipid vesicles serve as elementary models for biological membranes, enabling researchers to isolate and study complex mechanics of cell-like structures~\cite{dimova2019giant}. Consisting of bilayers of amphiphilic lipid molecules, vesicles are closed membranes of a few nanometers in thickness, dividing the aqueous interior from the surrounding aqueous exterior.  As most commonly studied lipids assembled to form the lamellae, their equilibrium shapes are well characterized by the Helfrich model, or its variants,~\cite{helfrich1973elastic,seifert1997configurations,deserno2015fluid}, which predominantly captures an elastic resistance to bending at fixed or preferred membrane area, but without resistance to in-plane shear.  As true biological membranes are inherently heterogeneous, composite structures comprising protein aggregates, phase-separated lipid ``raft'' domains, and polymerized cytoskeletal networks~\cite{simons1997functional,baumgart2003imaging,lipowsky2003domains}, considerable interest has focus on collective mechanics of multi-phase vesicles, possessing multiple 2D domains within the same vesicles.  The vast majority of studies have focused on scenarios where all domains are fluids but possess distinct mechanical properties (e.g. bending stiffness~\cite{julicher1993domain,julicher1996shape,baumgart2003imaging,baumgart2005membrane, ursell2009morphology,gutlederer2009polymorphism,semrau2009membrane,hu2011vesicles,aydin2014phase,waltmann2024patterning,sischka2025two} or spontaneous curvature~\cite{lipowsky1992budding,lipowsky1993domain,shimokawa2010phase,wolff2015budding,lipowsky2024multispherical,haselwandter2010minimal,haselwandter2011elastic, rower2023coarse}). The by far less understood scenario occurs in  multi-lipid vesicles in which the solidification of one of the two lipid phases is above room temperature~\cite{bandekar2012floret,chen2014large,beales2005solid,gordon2006lipid}, driving phase separation to composite vesicles in which 2D solid domains are integrated within a fluid membrane background, such as the example of a DOPC-DPPC vesicle shown in Fig. 1A.  Unlike fluid domains, solid domains exhibit an additional in-plane solid elasticity arising from a non-vanishing shear modulus.  

Notably, the 2D solid mechanics, when combined with the closed spherical topology of vesicles presents a basic mechanism of geometric frustration.  On one hand, bending energy alone (in absence of spontaneous curvatures) favors globally uniform spherical shapes~\cite{seifert1991shape}, with elastic energy increasing as the vesicles are deflated.  This is consistent with known elastic energy of the homogeneous fluid vesicle ground states shown in Fig. 2B, decreasing monotonically with reduced volume $\bar v=\sqrt{36\pi}V/A^{3/2}$ ($V$ and $A$ are the respective volume and area) to the minimal energy spherical shape at $\bar{v} \to 1$.  However, vesicles are nanometrically thin while lateral dimensions (i.e. diameter) exceed tens of microns for giant unilamellar vesicles (GUVs)~\cite{dimova2019giant}, suggesting that in-plane distortions required to impose non-zero Gaussian curvature on solid domains is under most conditions much more energetically expensive relative to bending elasticity.  Hence, it can be expected that solid domains expel Gaussian curvature which much be taken up by the surrounding fluid phase due to the net positive Gaussian curvature required by spherical topology and solid domains therefore non-spherical exhibit energy minimizing states at reduced volume $\bar{v}^* < 1$ (i.e. zero tension and internal pressure).  This characteristic behavior is shown in Fig. 2B for two models of solid domains:  a rigid, planar disc studied previously~\cite{jeon2024shape}; and a thin, circular elastic plate, which is the central focus of the current study.  While there are differences in the minimal-energy $\bar{v}_*$ between these two models, both show in Fig. 2C that presence of a solid domain leads to negative tension for deflated ($\bar{v} < \bar{v}_*$) and positive tension for inflated ($\bar{v} > \bar{v}_*$) regimes respectively.  

This article aims to address two overarching questions about the role of 2D solid elasticity in composite vesicles possessing single solid, yet flexible domains.  While solid domains can be formed with a variety of numbers and shapes, depending on crystalline phase, defects, and most significantly kinetics of solidification~\cite{lipowsky2003domains,beales2005solid,schneider2005shapes,chushak2005solid,gordon2006lipid,shimokawa2010phase,hamada2011lateral,vernizzi2011platonic,sknepnek2012buckling,chen2014large,wan2024flower,wan2024thermal,wan2026sculpting}, we focus on the elementary case of single circular 2D solid domain.  First, we ask how does the geometrically nonlinear resistance to Gaussian curvature effect the overall vesicle shape and mechanics?  In a previous study, we found that strictly rigid, but sufficiently large, solid domains profoundly shift the symmetries of optimal vesicles to largely oblate, relative to their predominating prolate shapes for homogeneous fluid vesicles~\cite{jeon2024shape}.  Here, we show that solid domain flexibility, notwithstanding its inextensibility, largely allows composite vesicles to maintain their predominately prolate shapes in the deflated regime.  This is facilitated by solid domains largely cylindrically bent ``rolls" that easily conform to preferred prolate geometry, without changes in Gaussian curvature.  Our second question focuses on the inflated regime, asking how geometric frustration encountered as a vesicle approaches increasingly spherical shapes impacts the morphology of solid domain?  For this we show that sufficiently thin (relative to the vesicle size) solids exhibit a characteristic sequence of elastic patterns extending from nearly-isometric rolls and folded shapes to non-isometric patterns of crumples and wrinkles prior to the axisymmetric spherical cap of the fully-inflated case.  We show that transitions between these patterns, as well as variation of their features, are controlled by the tension in the membrane compared to two characteristic mechanical scales, one set by bending energetics of the domain and the other by non-isometric deformations of 2D solid domain.  For the highly-inflated regime, we show that wrinkle patterns and transitions to crumpled and axisymmetric patterns follow a close correspondence to elastic sheets on fluid drops and are well-described quantitatively by the so-called far-from-threshold theory (FFT) of wrinkling~\cite{davidovitch2011prototypical,king2012elastic, paulsen2016curvature}.  We map out frustrated morphologies of solids as function of relative inextensibility of solid domains, which is effectively controlled by the vesicle size relative to nanoscale elastic thickness of the membrane, and argue that the complex sequence of isometric to non-isometric patterns intermediate to smooth rolls and caps is characteristic for GUVs, but disappears at smaller size ranges for so-called large unilamellar vesicles (LUVs) and smaller.

This article is organized as follows.  In Section \ref{sec: methods}, we first describe our composite fluid-solid vesicle elastic model and the Surface Evolver method used to implement it, and briefly introduce key parameters and physical scales that ultimately control equilibrium shapes.  We then begin in Section~\ref{sec: gs}, with a brief discussion of minimal energy (tension-free) shapes, comparing the distinct effects of solid inextensibilty (i.e. reduced elastic thickness)  versus solid to fluid bending stiffness ratio, while in the remainder of the study we consider fluid and solid domains to be equally stiff to bending.  We then consider the effect of solid domain elasticity on deflated vesicle shapes in Section~\ref{sec: def} before turning to frustrated elastic patterns in solid in inflated vesicles in Section~\ref{sec: inf}, characterizing their patterns as function of membrane tension, solid domain size and elastic thickness of membranes.  Finally, we conclude in Section~\ref{sec: conc}, presenting experimental evidence of shape transitions in inflated DOPC-DPPC composite GUVs, and discussing further experimental implications and open questions prompted by our results.  In addition to several Appendices detailing methods used, which include, in Appendix~\ref{app: bend}, additional results on the gross morphology dependence of composite fluid-solid vesicles on the solid-to-fluid bending stiffness ratio.

\section{Elastic energy of fluid-solid composite vesicles}
\label{sec: methods}

We consider an intrinsically flat, solid domain (i.e. defect-free crystalline) of circular shape of equilibrium (stress-free) area $A^0_\text{solid}$ integrated into a vesicle whose remaining area $A_\text{fluid}$ is a fluid membrane, enclosing at total volume $V$. For the fluid membrane, we consider a fixed area ensemble, such that elastic energy of the fluid membrane is governed entirely by the Helfrich bending energy~\cite{helfrich1973elastic}.  In addition to bending menchanics, the solid domain possesses both non-vanishing in-plane shear and area dilation modulus. This introduces an in-plane strain energy cost that strongly penalizes changes to the solid's 2D metric, and correspondingly changes of Gaussian curvature, as is well known feature of 2D plate mechanics~\cite{landau1986theory}. The bending energy for the entire composite vesicle and the strain energy for the solid domain are defined, respectively, as:
\begin{equation}
    E_\text{bend}=\frac{B_\text{f}}{2}\int_{\rm fluid} \d A\ (2H)^2+\frac{B_\text{s}}{2}\int_{\rm solid} \d A\ (2H)^2
\end{equation}
\begin{equation}
    E_\text{strain}=\frac{Y}{2(1+\nu)}\int_\text{solid}\d A\ \left[\text{Tr}(\varepsilon^2)+\frac{\nu}{1-\nu}(\text{Tr}~\varepsilon)^2\right]
\end{equation}
where $H$ and $K$ are the respective mean and Gaussian curvatures, with the bending response parameterized by the fluid and solid (mean curvature) bending moduli, $B_\text{f}$ and $B_\text{s}$, and we consider the case where deviations from tangential matching between the fluid and solid domains require divergent bending costs. For the solid domain, $\varepsilon_{ij}$ represents the in-plane 2D strain tensor, while $\nu$ and $Y$ are the 2D Poisson's ratio and 2D Young's strain modulus, respectively. The total elastic energy is the sum of these contributions
\begin{equation}
E_\text{el}=E_\text{bend}+E_\text{strain}
\end{equation}
We neglect bending terms linearly proportional $K$ for both domains, as these lead to constant contributions for the regime of interest, nearly-inextensible solid domains~\footnote{For isometric solid deformations, the Gauss-Bonnett theorem~\cite{kamien2002geometry}, along with tangential matching of fluid-solid domains leads to a shape-invariant contribution from linear terms in $K_{\rm fluid}$ due to the fixed genus of the vesicle.  Even, under extreme inflation where even the solid adopts Gaussian curvature, it is straightforward to show that $E_\text{bend} \ll E_\text{strain}$, so that the bending energetics (including linear terms in $K$) do not modify the gross morphology and elastic energy of vesicles}.


We work in an ensemble where the fluid area ($A_\text{fluid}$), the unstrained solid area ($A^0_{\rm solid}$), and the total enclosed volume ($V$) are held fixed.  While solid elasticity penalizes distortions of $A_{\rm solid} \neq A^0_{\rm solid}$, we achieve target fluid area and enclosed volume via minimization of the thermodynamic potential (free energy) 
\begin{equation}\label{eq: freeenergy}
    F=E_\text{el}+\gamma A_\text{fluid}-PV
\end{equation}
where membrane tension $\gamma$ and the pressure difference (inside relative to outside the vesicle) $P$ act as the Lagrange multipliers conjugate to the fluid area $A_\text{fluid}$ and the enclosed volume $V$, respectively~\cite{jeon2024shape}.  We perform these elastic ground state minimization via the finite-element framework, Surface Evolver (SE)~\cite{brakke1992surface, brakke2013surface}.  This approach optimizes the generalized elastic energy over vertex positions of triangulate meshes of fluid-solid composite vesicles.  Here, we perform studies on meshes including 88,320 triangles per vesicle, using methods previously implemented in~\cite{jeon2024shape}, and summarized in Appendix \ref{app: SE}.  

The elastic energy ground states are parameterized by two key geometric dimensionless quantities: the (target) solid area fraction $\Phi$ and the reduced volume $\bar{v}$:
\begin{equation}
    \Phi=\frac{A_\text{solid}}{A},\qquad
    \bar{v}=
    \frac{V}{\frac{4 \pi}{3} R^2} = \sqrt{36\pi}\frac{V}{A^{3/2}} 
\end{equation}
where $A=A_\text{fluid}+A^0_\text{solid}$ is the total vesicle surface area, $V$ is the enclosed volume, and $R=\sqrt{A/4\pi}$ is the characteristic size scale of the vesicle. Here, $\bar{v}$ measures the relative inflation of the vesicle compared to a perfect sphere ($\bar{v}_\text{sphere}=1$).  We consider the case of $\Phi \leq 0.15$, for which it is straightforward to show that area distortions of the solid are small, such that $A_\text{solid}\simeq A^0_\text{solid}$ and equilibrium and target values of solid fraction $\Phi$ are negligibly different.  

In addition to these geometric parameters, we define two key dimensionless parameters that quantify the solid domain mechanics,
\begin{equation}
    \bar{t}=\frac{t_{\rm s}}{R}= \sqrt{\frac{B_\text{s}}{Y}}\frac{1}{R},\qquad
    \beta=\frac{B_\text{s}}{B_\text{f}} .
\end{equation}
The first is a dimensionless measure of the ratio of bending to in-plane strain elasticity and the solid, where $t_{\rm s} \equiv \sqrt{B_\text{s} /Y}$, is a length scale which we call {\it elastic thickness} which is proportional to true thickness for the simplest models of homogeneous solid plates~\cite{landau1986theory,audoly2010elasticity}.  Here, on dimensional grounds and based on estimates of lipid bilayer membrane and area elasticity discussed later~\cite{rawicz2000effect} we expect $t_{\rm s}$ to be on the scale of few nanometer thickness of the membranes themselves, whereas $R$ is of order of 10s of microns for GUV scale vesicles, corresponding to dimensionless thickness that is asymptotically small $\bar{t} \sim ~ 10^{-4}$, which suggests that solid domains are relatively inextensible compared to their bendability in this regime.  The second parameter, $\beta$, characterizes the ratio of bending stiffness of solid to fluid domains, which is generally expected to be greater than one, due to increased order implied by solid relative to fluid state, although it is unlikely to be orders of magnitude larger unless the thickness of solid and fluid membrane phases is significantly different.



In the following we analyze the resulting equilibria through their localized curvature distributions and total elastic energies. Throughout, we scale lengths by powers of $R$ such that dimensionless mean and Gaussian curvatures rescaled by the vesicle size: $HR$ and $KR^2$.  We scale total elastic energy by the fluid bending stiffness $B_\text{f}$:
\begin{equation}
    \bar{E}=\frac{E_\text{el}}{8\pi B_\text{f}}
\end{equation}
where the prefactor of $8\pi$ normalizes to the bending energy of homogeneous spherical vesicle (i.e. $\bar{E} = 1$ for $\Phi \to 0$ and $\bar{v} \to 1$).  

Before proceeding, we briefly introduce two mechanical scales for in-plane tension $\gamma$.  The first corresponds to the mechanical scales required by bending of the vesicle,
\begin{equation}
\gamma_{\rm bend} \equiv \frac{B_\text{f}}{R^2}. 
\end{equation}
In Fig. 1, we find that reduced tension, normalized by this tension scale 
\begin{equation}
    \bar{\gamma} \equiv \gamma/\gamma_{\rm bend}
\end{equation} 
is of order $10^0$, indicating that bending mechanics dominates vesicle thermodynamics over most of reduced volume range, with important exception of divergence at high inflation when $\bar{v}$ is increased sufficiently above $\bar{v}^*$ such that $\bar{\gamma}  \gg 1$.  For much higher inflation, tensions grow to reach the mechanical scale set by solid strain mechanics.  To estimate this scale, we note that in-plane strain required to conform an elastic sheet of radius $W$ to sphere of curvature $1/R$ is $(W/R)^2$~\cite{grason2016perspective,meng2014elastic,schneider2005shapes}.  Since $W\propto \sqrt\Phi R$ for the composite vesicle, this implies the scale of 2D mechanical stress
\begin{equation}
\gamma_{\rm strain} \equiv Y \Phi. 
\end{equation}
When $\gamma \gtrsim \gamma_{\rm strain} $, 2D solid strain elasticity dominates the collective mechanics of the vesicle.  Using our definition of dimensionless thickness, we note that 
\begin{equation}
\gamma_{\rm strain} = \big( \Phi/\bar{t}^2 \big) \gamma_{\rm bend} , 
\end{equation}
and hence we expect $\gamma_{\rm strain} \gg \gamma_{\rm bend}$ for sufficiently large (and elastically thin) vesicles, such that these tension scales are well-separated.  A central aim of this study is to characterize the collective behavior that characterizes this crossover from bending-dominated to in-plane strain dominated mechanics. In this highly inflated regime, therefore, we introduce a second reduced tension, normalized by geometrically-induced in-plane stress of spherical confinement defined by 
\begin{equation}
    \tilde{\gamma} \equiv \gamma/\gamma_{\rm strain} =  \big( \bar{t}^2/\Phi \big)  \bar{\gamma} ,
\end{equation} 
and show that the fully inflated limit ($\bar{v} \to 1$) corresponds to the regime $\tilde{\gamma} \gtrsim 1$.

\section{Minimal-energy (pressure-free) states}\label{sec: gs}

\begin{figure*}[t!]
\begin{center}
    \includegraphics[width=0.75\textwidth]{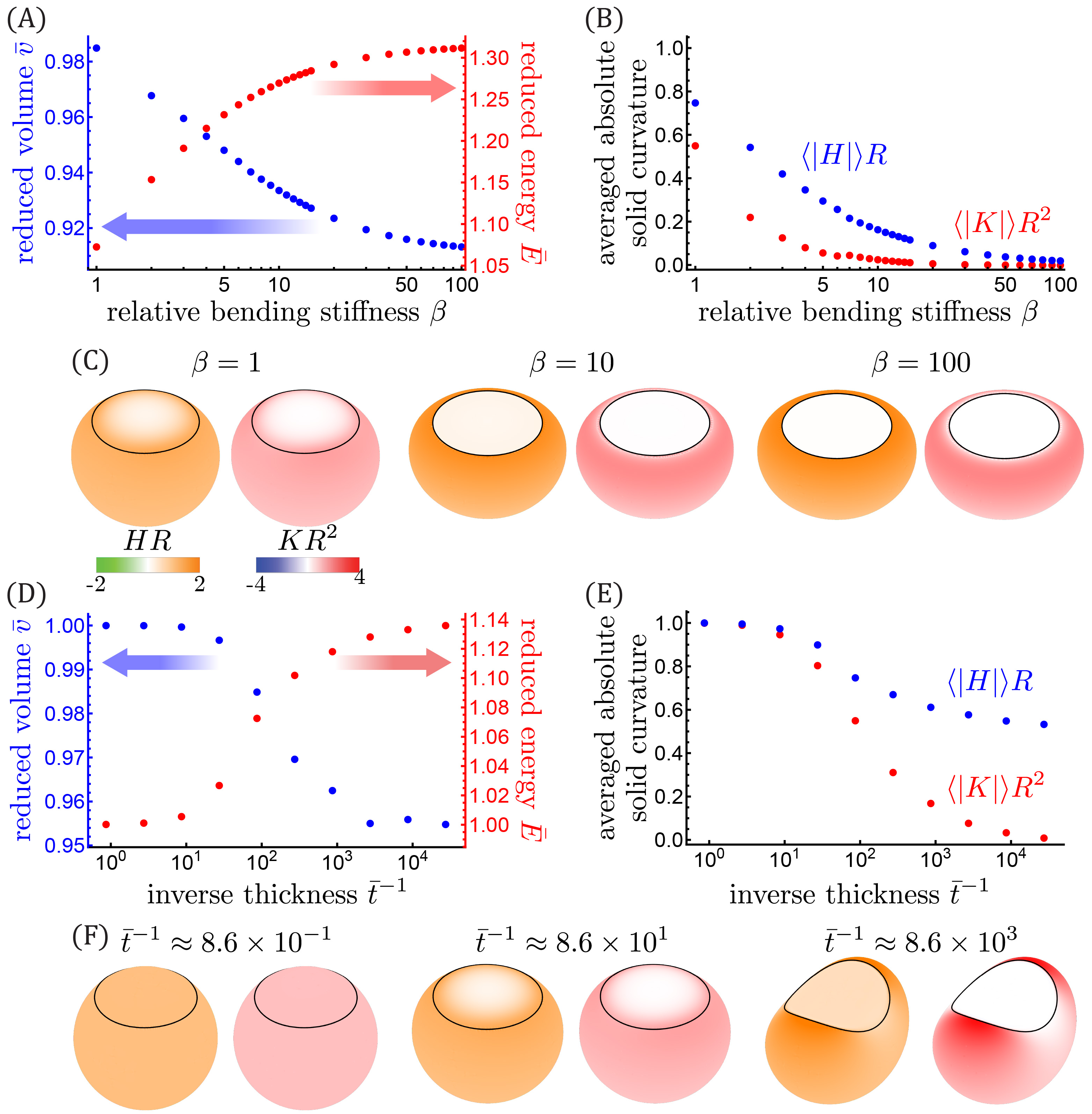}
\caption{\label{fig: zero_tension-pressure}\textbf{Roles of solid domain bending stiffness and inextensibility in ground state shapes} (A) Plots of reduced volume and reduced energy as functions of the relative bending modulus $\beta$ for $\Phi=0.15$ and $\bar{t}=1.2\times10^{-2}$ (i.e. extensible solid). (B) Averaged absolute mean ($|H|R$) and Gaussian ($|K|R^2$) curvatures of the solid domain for increasing $\beta$. (C) Example structures colored by local mean and Gaussian curvatures for plotted results in (A) and (B). (D) Reduced volume and reduced energy plotted versus inverse thickness $\bar{t}^{-1}$ for $\Phi=0.15$ and $\beta=1$ (i.e. equal fluid and solid bending stiffness), with averaged absolute mean and Gaussian curvatures of the solid domain in (E) and corresponding example ground states example structures colored by local mean and Gaussian curvatures in (F).}
\end{center}
\end{figure*}

For homogeneous fluid vesicles, the elastic energy is governed purely by bending, which favors globally uniform mean curvature states for closed topologies, i.e. minimum energy is achieved by perfectly spherical symmetry ($\bar{v}_*=1)$. Deflating homogeneous vesicles gives rise to a well-studied sequence of symmetry-breaking shape transformations, including prolate, oblate, and stomatocyte branches~\cite{seifert1991shape,ziherl2005nonaxisymmetric,reboucas2024stationary}. As shown in Fig. 1, fluid-solid composite vesicles exhibit minimal energy for non-spherical shapes ($\bar{v}^*<1$), leading to a non-monotonic dependence of elastic energy on reduced volume.  Before describing the equilibrium shapes in the deflated ($\bar{v}<\bar{v}^*$) and inflated ($\bar{v}^*<\bar{v}<1$) regimes, we first describe minimal energy shapes.  Strictly speaking, these states are minimized with respect to $V$ for fixed $A_{\rm solid}^0$ and $A_{\rm fluid}$, and are, hence, pressure-free states~\cite{seifert1997configurations}.\footnote{Under conditions where $E_{\rm strain} \to 0$, it is well known that the Helfrich bending energy is conformally invariant, meaning that it independent of scale $R$ and only a function of $\bar{v}$.  In that case, $P$ and $\gamma$ are both proportional to $\partial E_{\rm bend}/\partial \bar{v}$, implying that both pressure and membrane tension vanish at the same point.  And while in-plane strain may be strictly vanish in composite vesicles, we find that $E_{\rm strain}\ll E_{\rm bend}$ near to minimal energy states, such that we expect these states are very nearly tension-free, as well as pressure free.}

In a previous study~\cite{jeon2024shape} we considered the extreme case of strictly planar solid domain shapes, and found that with increasing solid fraction, the minimal-energy ($\bar{E}_*$) increases while the corresponding reduced volume ($\bar{v}_*$) decreases, while ground states remain generically planar and oblate.  In Fig.~\ref{fig: zero_tension-pressure}A-C, we consider flexible solid domains, first considering the case of reduced thickness fixed to modest value $\bar{t}=1.2\times 10^{-2}$ and fixed $\Phi = 0.15$, while varying the ratio of solid-to-fluid bending stiffness $\beta$.  These show that from comparable bending stiffness $\beta \approx 1$, and this relatively extensible value of reduced thickness, optimal vesicles are close to spherical inflation, corresponding to the significant Gaussian and mean curvature in the solid domain.  As $\beta$ increases, these optimal reduced volume and solid domain curvatures decrease towards zero.  The sequence of optimal morphologies in Fig.~\ref{fig: zero_tension-pressure}C show that ground state shapes remain axisymmetric, but that the shape and energetics only tend to saturate to values that are closely approximated by the rigid planar case when the bending stiffness contrast between fluid and solid is especially drastic (i.e. $\beta \gtrsim 50$).

We next consider, in Fig.~\ref{fig: zero_tension-pressure}D-F the case of equal-bending stiffness ($\beta =1$) but variable reduced thickness, ranging from highly extensible ($\bar{t} \approx 1$) to nearly inextensible ($\bar{t} \approx 10^{-5}$) solid strain energetics.  In this case, as shown in the morphology sequence in Fig.~\ref{fig: zero_tension-pressure}F, we find a transition from nearly spherical ground states, with highly curved cap shapes for the solid domains for large reduced thickness, to two-fold symmetric ground states at low $\bar{t}$.  Notably, ground state symmetry shifts from quasi-prolate to quasi-oblate around an threshold range of thickness $\bar{t}\approx 10^{-2}$.  Fig.~\ref{fig: zero_tension-pressure}E, shows that while this transition corresponds to expulsion of Gaussian curvature from the solid domain, there remains a large residual amount of mean curvature in the ground states of nearly inextensible composite vesicles.  The shape of solid domains in these ground states correspond to {\it rolls}, i.e. smoothly and cylindrical curved solids with fairly homogeneous (mean curvature) bending energy density and essentially vanishing in-plane strain.


Based on these results, we find that except under extreme conditions where solid domains are an order of magnitude (or more) stiffer to bending that fluid domains, they expel Gaussian curvature without maintaining appreciable mean curvature, meaning their shapes deviate, in general from perfectly planar.  Hence, in the remainder of this article we restrict our attention to equal bending stiffness ($\beta = 1$) in order to isolate the impact of the solid domain's defining feature, its resistance to in-plane strains, as quantified by variable dimensionless thickness $\bar{t}$.   In Appendix~\ref{app: bend}, we include additional results for the morphology and thermodynamics of composite vesicles for case of asymmetric bending stiffness ($\beta > 1$).

\section{Deflated vesicles}\label{sec: def}

As established in Section~\ref{sec: gs}, a non-vanishing shear elasticity shifts the ground state configuration away from the spherical limit (i.e. $\bar{v}^*<1$), due spatially inhomogeneous Gaussian curvature favored composite fluid-solid vesicles. This geometric frustration results in a non-monotonic dependence of energy on the reduced volume $\bar{v}$, splitting the phase space into a deflated regime and an inflated regime.  We now consider the effect of solid domain elasticity on the energetics and morphologies of the deflate region.

\begin{figure*}[t!]
\begin{center}
    \includegraphics[width=0.75\textwidth]{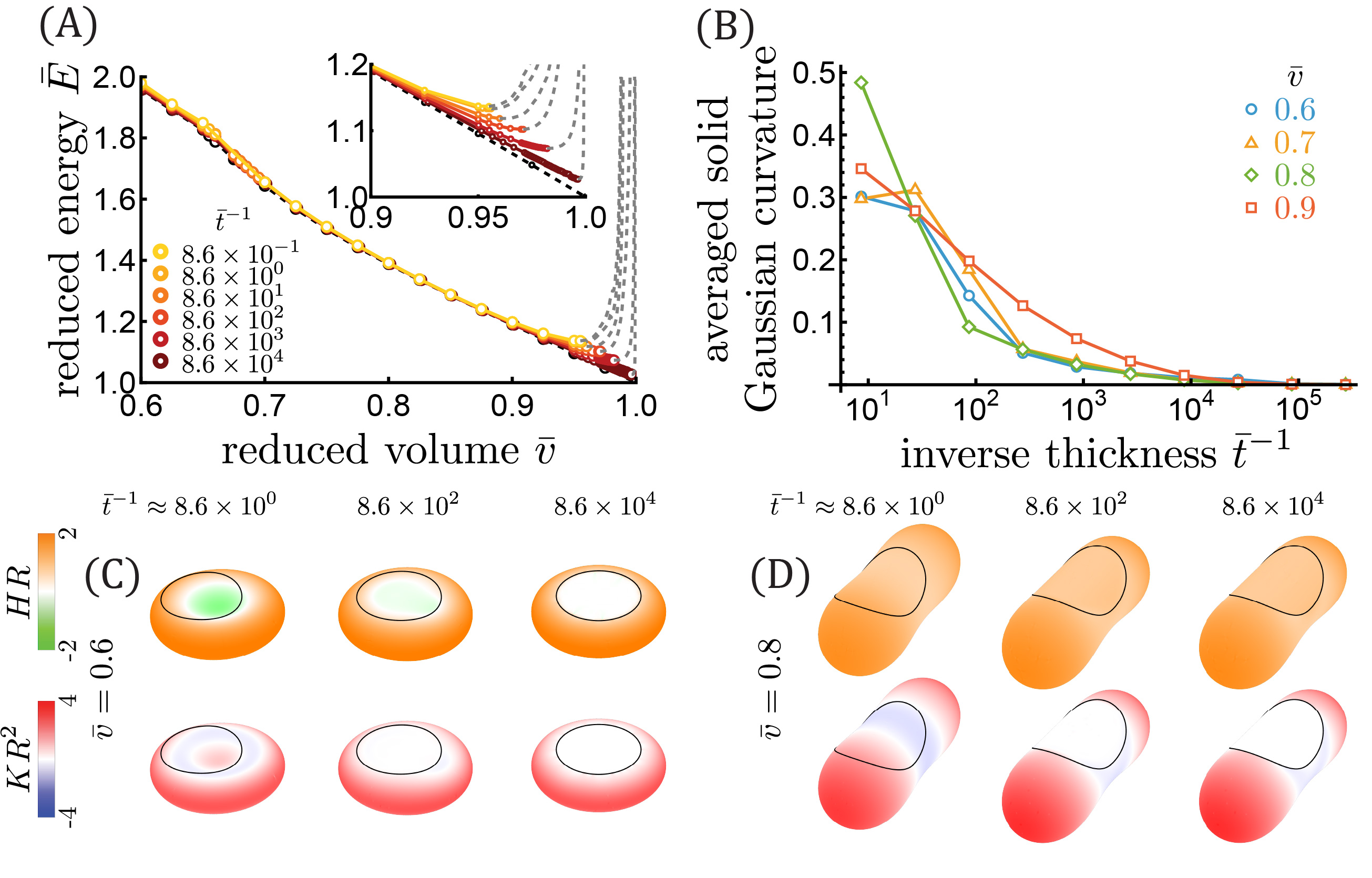}
\caption{\label{fig: deflated}\textbf{Solid domain elasticity has weak effect on deflated vesicle equilibria} (A) Reduced energy $\bar{E}$ plotted as a function of reduced volume $\bar{v}$ for various elastic thicknesses $\bar{t}$ at $\Phi=0.15$, colored regions highlighting the deflated (negative pressure) regime, while the inflated regimes is shown as dashed curves. (B) Averaged absolute Gaussian curvature of the solid domain $\langle|K|\rangle R^2$ as a function of the inverse thickness $\bar{t}^{-1}$, and correspond example ground states colored by local mean and Gaussian curvatures for $\bar{v}=0.6$ and $\bar{v}=0.8$ in (C) and (D), respectively.}
\end{center}
\end{figure*}

Figure~\ref{fig: deflated}(A) illustrates the energetics and morphology of composite vesicles, with fixed solid fraction $\Phi = 0.15$, across various elastic thicknesses ranging from extensible ($\bar{t}^{-1} \approx 1$) to relatively inextensible ($\bar{t}^{-1} \approx 10^5$), where the dashed portions of the curves correspond to the inflated regimes to be discussed in the next section. Figure~\ref{fig: deflated}(B) plots the averaged absolute Gaussian curvature over the solid domain as a function of the inverse thickness $\bar{t}^{-1}$, while Figures~\ref{fig: deflated}(C) and (D) provide example structures shaded by their spatial curvature distributions, showing both characteristic prolate and oblate equilibria. Like the case of minimal energy states, as the thickness decreases, the solid domain expels Gaussian curvature into the surrounding fluid background, resulting in a nearly vanishing absolute Gaussian curvature in the thin-sheet limit ($\bar{t}<10^{-3}$). Remarkably, despite the evident non-zero Gaussian curvature in for larger elastic thickness, the total elastic energy depends only weakly on the thickness and closely collapses onto the energy curve of homogeneous fluid vesicles.  That is, the shape and energetics of fluid-solid composite vesicles are insensitive to the inextensibility of 2D solid domain in the negative tension and pressure regime.

This independence of energetics on $\bar{t}$ of deflated vesicles can be understood by examining the localized structural adaptations. To mitigate the prohibitive penalty of in-plane strain, the solid domain actively localizes to regions where the underlying ground-state geometry naturally supports isometry, i.e. regions that are nearly zero $K$, or else require minimal perturbations of global shape to achieve $K\approx0$ at the location of solid domain. Specifically, the domain positions itself either on the relatively flat polar faces of an oblate vesicle, or on the cylindrical ``waist'' of a prolate vesicle---regions characterized by minimal Gaussian curvature but finite mean curvature. By occupying these optimal geometric niches, at inextensible solid conforms as a developable patch (e.g., a cylindrical or conical roll) that integrates smoothly into the global vesicle without significant perturbation of fluid membrane curvature, an effect that was predicted in a related model of thin sheet positioning on curved fluid interfaces~\cite{barakat2022curvature}. For this reason, thermodynamics of this fluid-solid composite closely track the morphology of a homogeneous fluid vesicle, thereby bypassing solid strain energetic cost almost entirely.  

Notably, as seen in the inset of Figure~\ref{fig: deflated}(A), energetics of the composite ultimately peel away from the homogeneous vesicle case as the minimal-energy state is approached, by an amount that increases for decreasing reduced thickness.  Additionally, changes of thickness below $\bar{t} \lesssim 10^{-5}$, do not seem to lead to further shifts in $\bar{E} (\bar{v})$ in this deflate regime, suggesting that these deflated and pressure-free results may be approaching a well-defined asymptotic limit of $\bar{t} \to 0$, governed by strictly isometric states (i.e. $E_{\rm strain} \to 0$) of the solid domain.  In the following section, we consider inflated states beyond this point, focusing foremost on elastically thin solid domains, to understand the extent to which nearly isometric states are favored and when inflation gives rise states of non-zero solid strain.

\section{Inflated vesicles}\label{sec: inf}

\begin{figure*}[t!]
\begin{center}
    \includegraphics[width=0.75\textwidth]{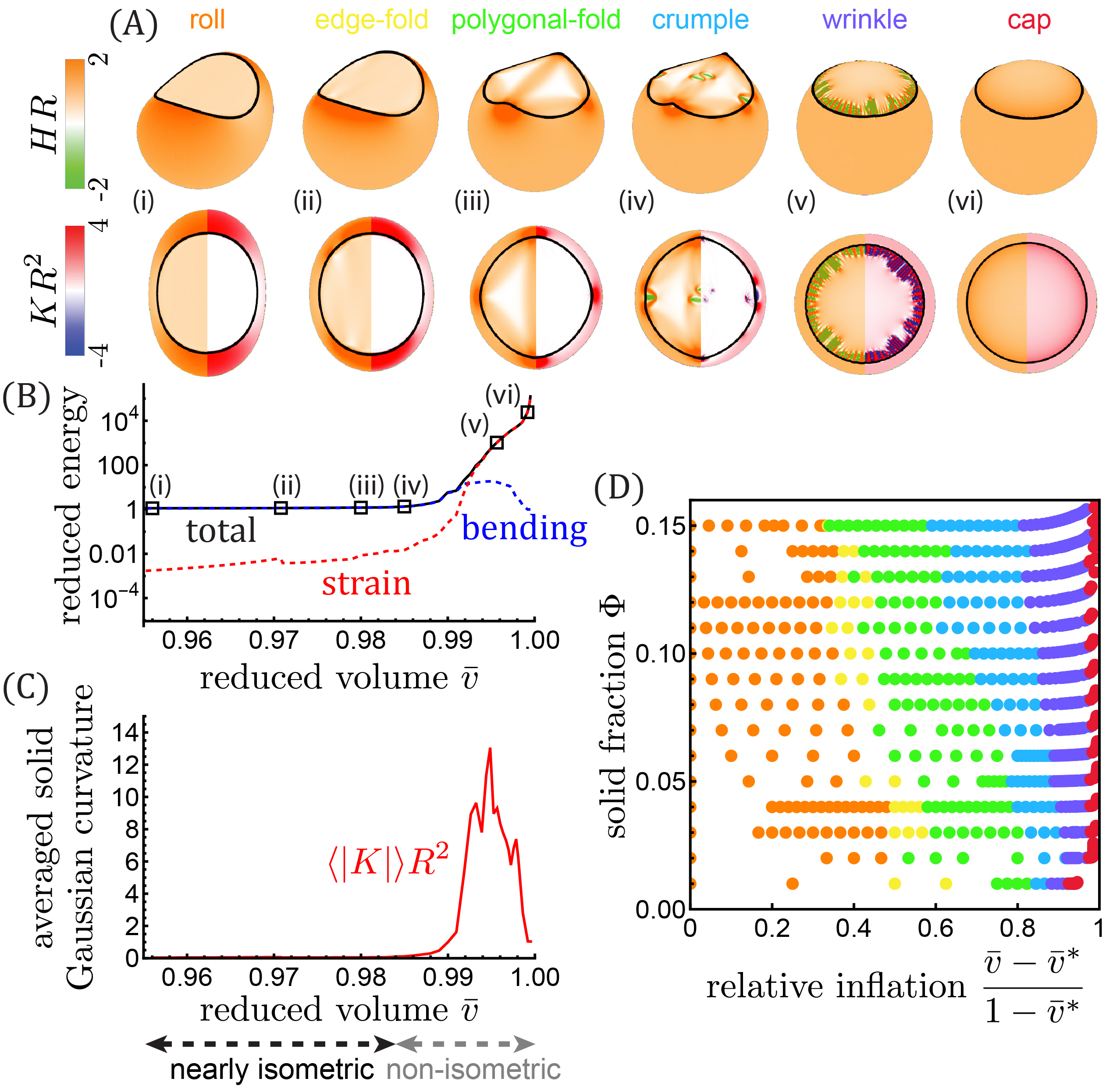}
\caption{\label{fig: inflated}\textbf{Sequence of elastic patterns on solid domains for increasing inflation.} (A) Six solid elastic patterns (roll, edge-fold, polygonal-fold, crumple, wrinkle, and cap) of inflated vesicles colored by local mean and Gaussian curvatures for an increasing sequence of reduced volume for $\Phi=0.15$ and $\bar t\approx1.2\times10^{-4}$ (i.e. relatively inextensible solid).  (B) Reduced energy (total, bending, and strain components) as a function of reduced volume $\bar{v}$ for $\Phi=0.15$ and $\bar t\approx1.2\times10^{-4}$, with labeled points (i)-(vi) corresponding to images in (A), and averaged absolute Gaussian curvature of the solid domain as a function of reduced volume $\bar{v}$ shown (C). (D) Phase diagram of the solid domain patterns ($\bar t\approx1.2\times10^{-4}$) as a function of solid fraction $\Phi$ and measure of {\it relative inflation}, $(\bar{v}-\bar{v}^*)/(1-\bar{v}^*)$.}
\end{center}
\end{figure*}

As described in the previous section, in the limit of vanishing thickness and zero pressure (tension) simple cylindrical bending of solid domains is favored to relax the combined fluid-solid bending energy without imposition of solid strain. However, as the vesicle is inflated, this uniaxial bending mode runs up against two fundamental constraints that frustrate isometrically rolled solid shapes. First, the Gauss-Bonnet theorem guarantees that the integrated Gaussian curvature of a closed vesicle is a constant $4\pi$. If the solid domain strictly maintains isometry ($K = 0$), the fluid membrane is forced to accommodate all of this excess Gaussian curvature. While this is apparently energetically inexpensive in the deflated regime due to the prevalence of regions of $K \approx 0$ on bending favored states where the solid domain can localize.  As pressure grows, the fluid itself becomes increasingly spherical, and nevertheless, must conform to match the solid domain, in an increasingly narrow and highly-curved boundary region between the two, an effect previously studied for the case of rigid solid domains~\cite{jeon2024shape}.  Second, for at least partial extensible (i.e. non-vanishing but small $\bar{t}$) solids, large enough pressures can eventually deform them to non-isometric states, even ultimately driving to volume maximizing spherical geometries.  We can anticipate that such non-isometric deformations should be encountered at pressure scale $P \gtrsim P_{\rm strain}= \gamma_{\rm strain}/R$.  

Here, we show, focusing primary on thin-solid limit ($\bar{t} \lesssim 10^{-3}$), that bridging between smooth rolls at zero pressure and axisymmetric, fully-inflated spherical caps at $\bar{v} \to 1$, solid domains exhibit a generic sequence of symmetry-breaking equilibria shown schematically in Figure~\ref{fig: inflated}A for the case of $\Phi = 15$:  rolls, edge-folded, polygonal-folded, crumpled, wrinkled and smooth caps.    The corresponding reduced elastic energy, in components from $E_{\rm bend}$ and $E_{\rm strain}$, are plotted from $\bar{v}^*<\bar{v}<1$ in Figure~\ref{fig: inflated}B, while Figure~\ref{fig: inflated}C shows the averaged absolute Gaussian curvature of the solid.  

We describe the detailed characteristics of each pattern in the next sub-sections, but summarize first their sequence with increasing inflation and their broad categorization on isometric (i.e. strain-free) or non-isometric states.  For this example, up to $\bar{v} \approx 0.985$, the solid domain successfully maintains a nearly isometric state ($K \approx 0$), and the total energy is heavily dominated by bending. In this regime, the optimal solid pattern transitions from smooth rolls to two-classes of nearly isometrically folded states, which we dub edge-folds and polygonal-folds.  However, as reduced volume increases further $\bar{v}\gtrsim 0.985$, elastic ground states crossover to clearly non-isometric states. This is evident from both the obvious growth to non-zero Gaussian curvature in the solid shown in Figure~\ref{fig: inflated}C, as well as the rapid upturn in $E_{\rm strain}$ and eventual overtaking of $E_{\rm bend}$ in the elastic energy.  To relax this mounting geometrically induced stress of spherical inflation, the solid evolves from folded states by adopting a patterns of localized elastic structures, crumples and radial wrinkles, until the vesicle is ultimately forced into a spherical, fully strained shape at maximum inflation ($\bar{v}=1$). 

Before turning to the mechanical parameters that select between these distinct parameters in the next sections, we first overview the elastic patterns of solids as functions to geometrical parameters, solid fraction $\Phi$ and relative inflation $(\bar{v} - \bar{v}^*)/(1 - \bar{v}^*)$, which measure excess reduced volume relative to the zero pressure ground state and normalized by inflation to the spherical state.  We find that, at least for this small $\bar{t}$, the sequence of patterns is maintained over this range of $\Phi \leq 0.15$, albeit with some shifts in transition points, whose origins we understand in the following sections.


\subsection{Modest inflation: nearly-isometric solid domain}

\begin{figure*}[t!]
\begin{center}
    \includegraphics[width=0.75\textwidth]{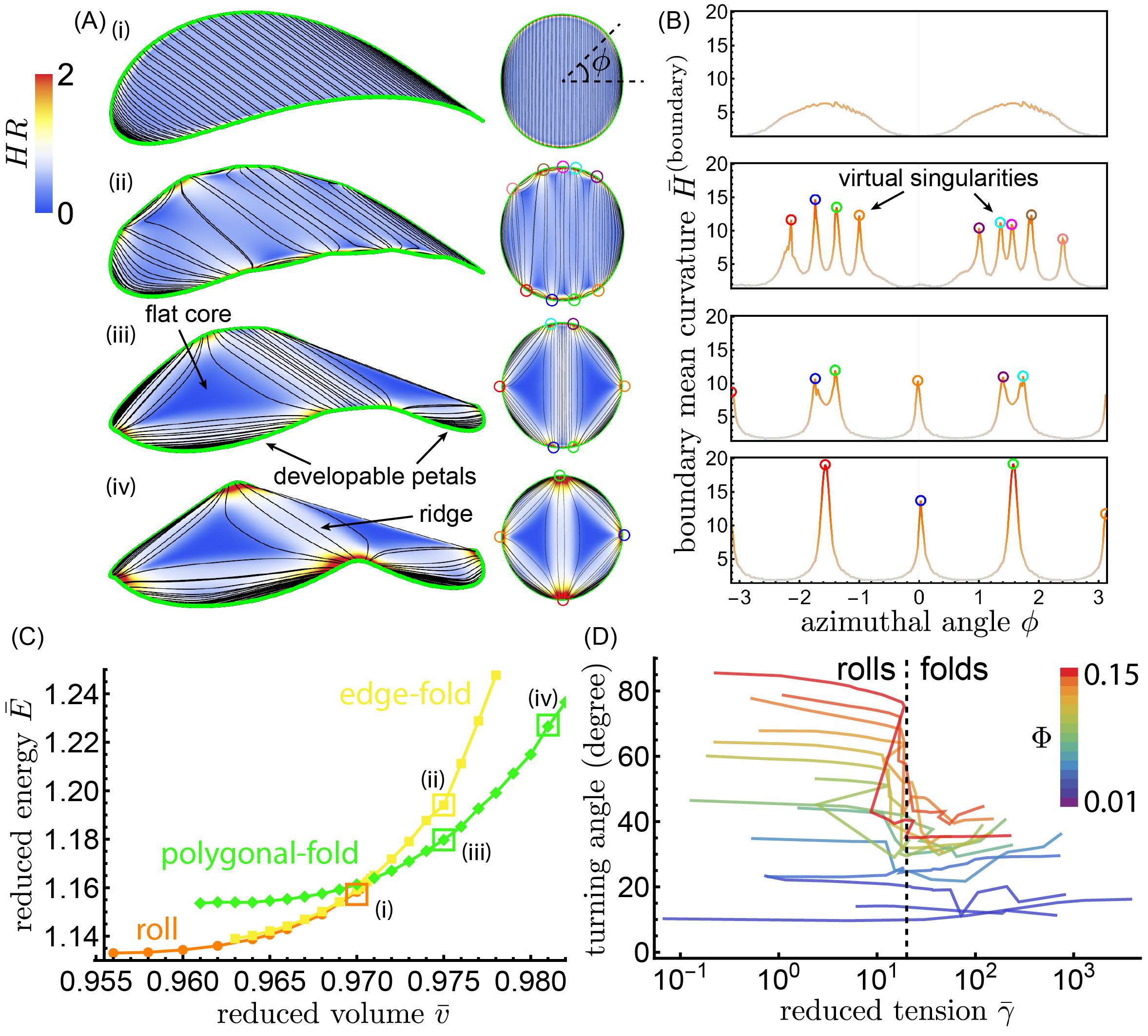}
\caption{\label{fig: isometry}\textbf{Developable folding of nearly isometric states.} (A) Solid domains colored by local mean curvature for three characteristic (nearly zero strain) patterns -- roll, edge-fold, and polygonal-fold -- and with streamlines along the least principal curvature direction. (B) Corresponding local mean curvature domain edge (on the fluid membrane side) as a function of the azimuthal angle $\phi$, showing peaked distributions for virtual singularities corresponding to the near intersection generator-like fold directions in the solid. (C) Reduced energy of 3 elastic patterns  over the range of $\bar{v}$ where they compete. Results in (A)-(C) correspond to $\Phi=0.15$, $\bar t\approx1.2\times10^{-4}$. (D) total turning angle over the solid domain along the greatest net bending direction as a function of reduced tension $\bar \gamma$, indicating a discontinuous change at a nearly constant $\bar{\gamma}$ threshold for fixed $\bar t\approx1.2\times10^{-4}$.  }
\end{center}
\end{figure*}

At modest inflation levels ($\bar{v} \gtrsim \bar{v}^*$), the solid domain maintains its nearly isometric character, ensuring that the total elastic energy remains governed primarily by bending.  As inflation is increased the smoothly rolled solid shape becomes unstable to additional nearly-isometric states, which maintain $E_{\rm stain} \approx 0$ while facilitating greater volume and negotiate optimal compromise between solid and fluid bending energy.  Hence, we expect that morphologies in this regime is controlled by the bending reduced tension, $\bar{\gamma} = \gamma R^2/B$.


In Figure~\ref{fig: isometry}A (for $\Phi=0.15$ and $\bar{t}^{-1}\approx 8.6\times 10^{3}$), we illustrate the three characteristic elastic patterns encountered with increasing inflation:  rolls, edge-folds and polygonal-folds (in two geometric variants).   In the roll phase, the streamlines of smallest principal curvature remain straight and parallel, acting as the continuous generators of cylinder, and mean curvature is fairly homogeneously distributed.  However, as the vesicle becomes more spherical, a global cylindrical bend becomes geometrically incompatible, leading to a 2-fold modulation fluid curvature required to meet the solid domain along respectively curved and straight tangent directions (i.e. the fluid membrane has to turn more away from the straight direction of the solid to meet the quasi-spherical bulk of the vesicle).  This profile of fluid membrane curvature at the domain edge (i.e. just outside the solid) is plotted in Figure~\ref{fig: isometry}B, and for the roll shows a smooth, spatially distributed curvature variation.

As inflation further increases, this smoothly bent shape of the solid and fluid composite become unstable to folded states.  As a first approximation, the solid remains isometric, and hence it can only be folded in certain way without introducing metric distortions.  These configurations are known as {\it developable surfaces}, and correspond to configurations where the surface is everywhere conically or cylindrically bent around straight lines, known as {\it generators}~\cite{witten2007stress}, which cannot cross.  We envisage this nearly-isometric folding in Figure~\ref{fig: isometry}A the streamlines of smallest principal curvature~\footnote{Note that streamlines are not perfectly straight, particularly at the edge of the solid domain, where  boundary layers exist for finite $\bar{t}$ leading to narrow regions of non-vanishing $K$}.  Notably, we find these streamlines converge towards point-like regions on on the boundary of high curvature.  Such configurations of near crossing of generator lines are known as ``virtual singularities'' as they indicate singular conical points just outside the solid boundary implies by the intersection of two fold lines, and they have been observed in frustrated states of isometric ribbons or else so-called capillary origami geometries of sheets wrapping drops~\cite{starostin2007shape,py2007capillary}.  Near intersection of fold directions lead to point-like concentration of curvature just outside the solid edge, which we detect at analysis in Figure~\ref{fig: isometry}B.

As shown in Fig.~\ref{fig: isometry}C, the roll first becomes unstable to a branch of edge-folded states, characterized by several virtual singularities on two opposing edges of the solid, with small edge folded ``petals'' bent down toward the fluid membrane, better compatibilizing it with the increasingly spherical shape of the fluid.  In the center of the solid, the curvature directions are noticeable no longer parallel (roughly zig-zagging between singular points at the edge), indicating modulated bending throughout the solid.  This edge-folded state is by comparison less organized than the subsequent polygonally-folded state, and it is energetically favored in only a very narrow inflation range.

The predominantly stable folded states maintain the general two-fold symmetry favored by prolate geometries (and rolls), and are composed of either a hexagonal or square skeleton of folds at the exterior, and two or one fold domain the middle dividing between two triangular flat panels.  Notably the double-folded ridge of hexagonal pattern eventually merges to the single fold above a threshold inflation.

In Fig.~\ref{fig: isometry}D we plot turning angle in this modest inflation regime as function of bending reduced tension $\bar\gamma$.  Notably, the transition point from rolls to folded geometries is observed to occur at a roughly constant value of $\bar{\gamma} \approx 20$, consistent with the notion that these nearly-isometric states are governed purely by bending energetics.  Notably, in the next section, we consider the regime where $\bar{\gamma} \gg 10^3$ in which solid states crossover to non-isometric states.


\subsection{Strong inflation: non-isometric solid domain}

\begin{figure*}[t!]
\begin{center}
    \includegraphics[width=0.75\textwidth]{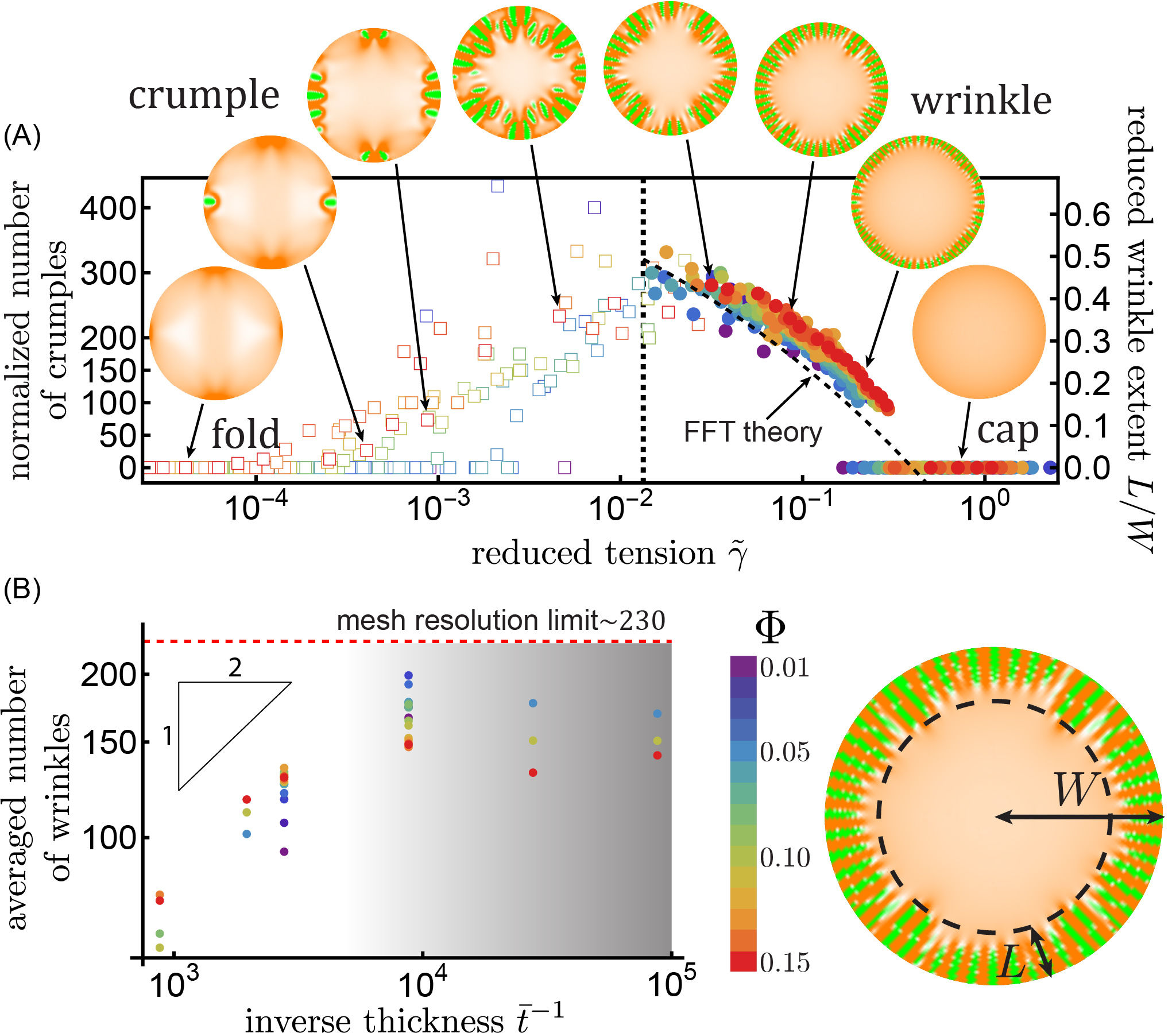}
\caption{\label{fig: non-isometry}\textbf{Crumpled and wrinkled patterns in highly-inflated vesicles.} (A) Number of crumples normalized by the solid fraction $\Phi$ (left axis, hollow squares) and the relative wrinkle zone extent $L/W$ (right axis, filled circles) as functions of the reduced tension $\tilde{\gamma}$. Points are colored according to the solid fraction as in the legend.  For $\Phi > 0.03$ plotted results in (A) correspond to elastic thickness $\bar{t} = 3.7 \times 10^{-4}$, while we show results for thinner solids $\bar{t} = 1.2 \times 10^{-4}$ for $\Phi \leq 0.03$ to main the far-from-threshold (FFT) wrinkling limit~\cite{king2012elastic}. The dashed curve denotes the scaling prediction for the wrinkle zone based on FFT, eq. (\ref{eq: L}).   (B) Averaged number of wrinkles plotted as a function of the inverse thickness $\bar{t}^{-1}$, showing the expected $\bar{t}^{-1/2}$ scaling for smaller wrinkle numbers and then saturating for larger $\bar{t}^{-1}$, consistent with mesh resolution limits (the horizontal dashed line shows an estimate for mesh-limited maximal number of wrinkles).  }
\end{center}
\end{figure*}

As inflation proceeds to extreme, spherical values ($\bar{v} \lesssim 1$), maintaining a nearly isometric solid becomes geometrically impossible and the solid develops strain.  The mechanisms for thin 2D solids to adopt non-zero Guassian curvature are particular complex, in large part due to the generic instability of thin solids to compression.  Depending on boundary loading, axisymmetric configuration of inflated solid plates (e.g. pressurized circular plates) acquire spatial regions of both compression and tension, and for sufficiently thin plates, such compressive zones are unstable to out-of-plane modes, leading to complex, spatially non-uniformed buckling patterns.  Here we show for inflated composite vesicles, this transition occurs first through a incorporation of increasing numbers of localized crumples, which eventually consolidate into organized radial wrinkles.  Indeed, this transition closely mirrors the morphologies of thin circular sheets on fluid droplets, where essentially the same crumple-to-wrinkle transition has been observed~\cite{king2012elastic,timounay2020crumples}.   Accordingly, we similarly anticipate the bending energetics to be subdominant relative to the solid strain energetics, and therefore, consider the evolution of the (sufficiently thin) solid patterns to be governed by strain reduced tension, $\tilde{\gamma} = \gamma/ Y \Phi$.  Notably, $\gamma$ is inversely proportional to {\it confinement parameter} defined in refs.~\cite{king2012elastic} and shown to govern the large-scale morphological features of floating thin sheets. 

Figure~\ref{fig: non-isometry}(A) shows the evolution of solid morphology from polygonal-folds to axisymmetric caps as function of strain reduced tension $\tilde{\gamma}$, plotting morphologies features associated with corresponding crumple and wrinkle patterns.  Nearly isometric folds eventually become unstable to crumple formation, above a tension we show below to be sensitive to thickness.  Crumples are localized regions that invert the deflection of the sheet from its surrounding~\cite{timounay2020crumples}.  If in the interior of the sheet, crumples are elongated and elliptical, and lead to effectively quadrupolar and localized distributions of Gaussian curvature associated with their ridge-like tips.  Here, we find ``half-crumples" that tend to localize at the boundary, with their long axis aligned with the radial direction.  This configuration may indeed be favored as means to maintain a more uniform ``latitude" for the solid edge.  We plot the scaled number density of crumples (i.e. total number divided by $\Phi$) with $\tilde{\gamma}$ for a range of solid fractions.  While the formation mechanism of crumples has remained unresolved, one consistent feature is their consolidation and transition to a wrinkle state, in which multiple crumples seemingly merge into extended wrinkles.  This is consistent with observed sequences in Figure~\ref{fig: non-isometry}(A), that show crumples increasing in number and localizing to an outer radial zone that eventually transforms to wrinkles  For elastic sheets on fluid droplets, this the crumple-to-wrinkle transition was phenomenologically observed at a critical range of large confinement.  Here, we find that the crumple-to-wrinkle transition corresponds to $\tilde{\gamma} \approx 10^{-2}$.  


Following the crumple phase, the solid domain transitions into a radial wrinkle state, with short-wavelength wrinkles extending from a finite radius $W-L$ from the sheet center to the sheets edge, smaller to the radial wrinkle morphology studied previous in spherical confined sheets, most specifically circular sheets on fluid drops)~\cite{king2012elastic}.  In that scenario, and in our case where bending energetics are negligible, the solid is subjected to an internal pressure $P \approx \gamma/R $ favors spherical bulging of its profile, while at the same time the surrounding phase (here, the fluid membrane) applies a lateral tension $\gamma$.  The spherical bulging tends to drive compression in the hoop in the outer radial zones of the solid, and wrinkles form as means to ``collapse'' the unstable compressive stress and relieve the ``excess material lengths'' in this region of length $L$.  As detailed in Appendix~\ref{app: wrinkle}, an asymptotic far-from-threshold (FFT) theory developed in refs.~\cite{davidovitch2011prototypical, king2012elastic, paulsen2016curvature} shows the size of this collapse zone is independent of solid thickness in the limit of vanishing $\bar{t}$ and is only a function of $\tilde{\gamma}$.  Specifically, the length of wrinkled zone is predicted to follow
\begin{equation}
\label{eq: L}
L = W\big[1 - (\tilde{\gamma}/\tilde{\gamma}_c)^{1/5} \big],
\end{equation}
where $\tilde{\gamma}_c \simeq 0.4$ is predicted to be a critical value of tension above which wrinkles are expelled from the solid (i.e. their length vanishes).  Figure~\ref{fig: non-isometry}(A) plots the observed length of the wrinkle zone as function of reduced tension on composite vesicles, showing remarkable fit-free agreement with the predictions of FFT theory and SE results for a range of solid domain fractions.  

While the gross features of the wrinkled morphology are independent of bending energetics, its fine features, specifically the wrinkle wavelength is controlled by the subdominant effects of bending.  Accordingly, the wrinkle wavelength $\lambda$ is expected to follow a characteristic proportionality $\lambda \propto (B/Y)^{1/4} \propto \bar{t}^{1/2}.$  The dependence of wrinkling on the inverse thickness $\bar{t}^{-1}$ is plotted in Figure~\ref{fig: non-isometry}(B). As expected, the number of wrinkles grows as $W/\lambda \sim \bar{t}^{-1/2}$ until it eventually saturates at the upper end of our inverse thickness range ($\bar{t}^{-1}\sim 10^4$). This saturation is an artifact of the finite-resolution of our triangulated meshes, where the fine wrinkling wavelength reaches the discrete spatial resolution limit of our finite element mesh (see Appendix~\ref{app: mesh_resolution} for a discussion of this bound). 

Because wrinkles vanish for $\gamma \gtrsim \gamma_{\rm axi} = \tilde{\gamma}_c Y \Phi$ this range of tension marks to the transition from spatially complex elastic patterns in the solid domain to smooth, axisymmetric, spherical cap morphologies of the solid domain that facilitate maximally inflated vesicles (i.e. $\bar{v} \to 1$)

\subsection{2D solid, elastic pattern selection }

\begin{figure*}[t!]
\begin{center}
    \includegraphics[width=0.75\textwidth]{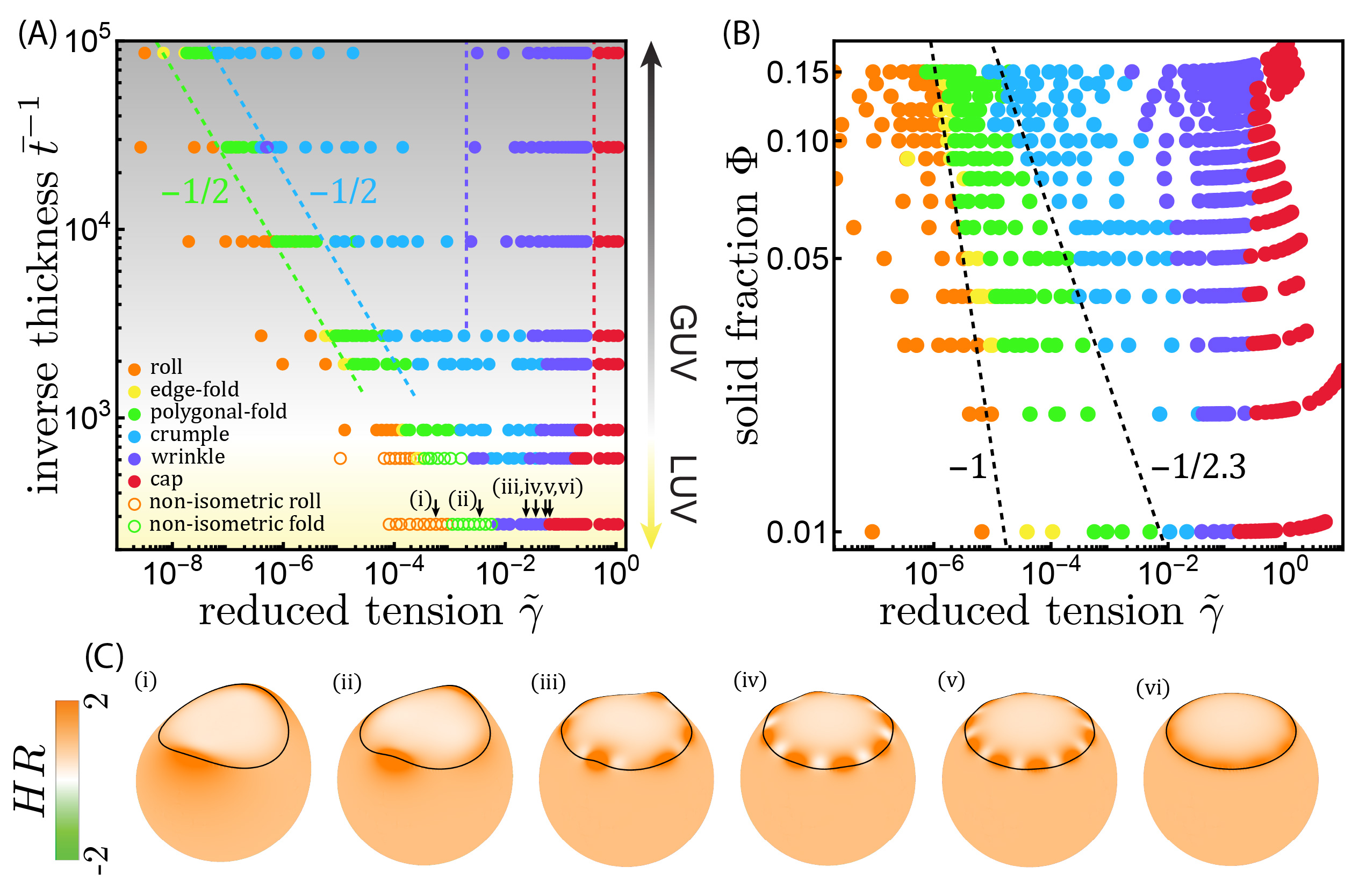}
\caption{\label{fig: phase} \textbf{Phase diagram of the solid domain patterns for inflated vesicles} (A) 2D $\bar{t}$-$\tilde{\gamma}$ phase map at constant $\Phi = 0.15$. The dashed lines highlight apparent transitions between (green) roll-fold, (blue) fold-crumple, (purple) crumple-wrinkle, and (red) wrinkle-cap, slopes giving estimate scaling dependencies, with the right-vertical axis labeling the estimated range of large- versus giant-unilamellar vesicles (assuming a fixed few nm elastic thickness). Open symbols denote two near-threshold patterns observed at low inverse thickness (i.e. extensible solids), non-isometric rolls and non-isometric folds, which are characterized by non-zero strain, shown for corresponding examples in (C) shaded by curvature.  (B) 2D $\Phi$-$\tilde{\gamma}$ phase map at fixed (small) reduced thickness $\bar t\approx1.2\times 10^{-4}$.  }
\end{center}
\end{figure*}

In the prior subsections, we found that elastic patterns of solid domains are controlled by three dimensionless parameters, reduced tension, solid fraction and ratio of thickness to vesicle size.  In Figure~\ref{fig: phase}, we plot phase diagrams of the solid domain elastic patterns in two 2D cuts, for fixed solid fraction $(\Phi = 0.15)$ in (A) and fixed scaled thickness $(\bar{t} \approx 1.2\times10^{-4})$ in (B).  Turning first to pattern selection in $\tilde{\gamma} - \bar{t}^{-1}$ plane, shown Figure~\ref{fig: phase}A, for sufficiently thin sheets $\bar{t}^{-1} \gtrsim 10^{3}$ we find a conserved morphological transition sequence: from rolls to edge-fold to polygonal-folds to crumples to wrinkles to caps. Notably, in this regime the transitions from crumples to wrinkles and then from wrinkles to caps are controlled only by $\tilde{\gamma}$ and {\it independent} of thickness $\bar{t}$, consistent with the previous studies of thin floating sheets~\cite{king2012elastic, timounay2020crumples} and the expectation that dominant energetics governing pattern selection in this strong-inflation (non-isometric) regime are independent of bending elasticity. In contrast, we observe that the threshold strain reduced tension for the roll-to-fold and fold-to-crumple transitions exhibit an evident $\tilde{\gamma} \sim \bar{t}^2$ scaling, consistent with these transitions set by the bending tension scale $\gamma_{\rm bend}=B/R^2$ since $\gamma_{\rm bend}/\gamma_{\rm strain} \propto \bar{t}^2$ and the idea that pure bending energetics dominates in the nearly-isometric states.  We note that, however, while it reasonable to expect roll-to-fold transition to be governed purely bending energetics (and set by tension scale $\gamma_{\rm bend}$), it is not clear why the onset to crumpling should be purely bending controlled, as the energetics of crumples and their behavior in the high-confinement limit are not currently understood~\cite{timounay2020crumples}.  

The consequence of this $\bar{t}$ dependence is that roll-to-fold and wrinkle-to-cap transition thresholds eventually converge for sufficient small inverse thickness, $\bar{t}^{-1} \approx 10^{3}$, such that all the intermediate elastic patterns between rolls and axisymmetric caps are largely squeezed out for $\bar{t}^{-1} \lesssim 10^{2}$.  Indeed, for the smallest inverse thickness, we find a nearly direct transitions from rolls to a narrow window of near-threshold wrinkling before axisymmetric caps, as well as clearly non-isometric (i.e. significantly strained) rolls and fold states~\footnote{Near-threshold wrinkling behavior is evident from apparent constant extent of wrinkling, while the wrinkle amplitude and wavelength decrease with increasing tension~\cite{davidovitch2011prototypical}}.  As $\bar{t}=t/R$ is a measure of elastic thickness relative to vesicle size, this suggests that the emergence of complex elastic patterns on solids (including e.g. folded and crumpled states) is strongly sensitive to vesicle size.   Estimating $t \approx 4~\text{nm}$ to be fixed, we estimate the typical range of GUVs versus LUVs on the vertical axis of Figure~\ref{fig: phase}A, suggesting that complex elastic patterns of solids are more likely to be observed in GUVs of multi-micron radius or larger, and are a far more modest tension scale.  

In Figure~\ref{fig: phase}B, we turn to elastic pattern selection as function of $\Phi$ and $\tilde{\gamma}$ at fixed large inverse thickness $\bar{t}^{-1}\approx8.6\times10^3$, corresponding to the aforementioned GUV range.  We find again, that the crumple-to-wrinkle and wrinkle-to-cap transitions values of $\tilde{\gamma}$ are independent of $\Phi$ consistent with FFT reasoning described previously from floating thin sheets~\cite{king2012elastic}.  In contrast, we find that the roll-to-fold transition exhibits a roughly $\tilde{\gamma} \sim \Phi^{-1}$, which would be consistent with the idea that threshold tension for transitions from smoothly curved to inhomogeneously folded, yet isometric, states occurs when tension reaches a threshold fraction of $B/R^2$, driving redistribution of bending energy between the fluid and solid membranes portions.  Notably, we find a stronger $\Phi$-dependence of the threshold between folded and crumple states $\tilde{\gamma} \sim \Phi^{-2.3}$, which is close to a threshold determined by a slightly elevated bending tension scale $B/W^2=\Phi^{-1} B/R^2$ by the solid domain size.  Hence, this evident difference in $\Phi$-dependence leads to a respective narrowing and widening of the windows of isometrically folded and non-isometric crumpled states as solid domain fraction grows.


\section{Discussion}\label{sec: conc}

In this article, we studied the morphological landscape of fluid-solid composite vesicles, highlighting how the geometric frustration between in-plane shear elasticity and global membrane bending elasticity governs equilibrium shape and naturally splits emergent morphological behavior into two, inflated versus deflated regimes. While deflated vesicles are largely insensitive to solid domain elasticity relative to homogeneous fluid bending elasticity, the inflated regime exhibits a rich spectrum of inhomogenously patterned shapes without counterpart to the pure fluid elasticity case. We find further, that solid domain patterns in the inflated regime are split into (nearly) isometric and non-isometric regimes, and accordingly transitions between patterns in these two regimes are governed by two mechanically distinct scales of membrane tension $\gamma_{\rm bend} = B/R^2$ and $\gamma_{\rm strain} = Y \Phi$.  For the GUV scale where most current experimental observations of fluid-solid composites exist, these two characteristic tension scales are separated by orders of magnitude, which provides the basis for the experimental implications of our findings.

\begin{figure}[t!]
\begin{center}
    \includegraphics[width=\columnwidth]{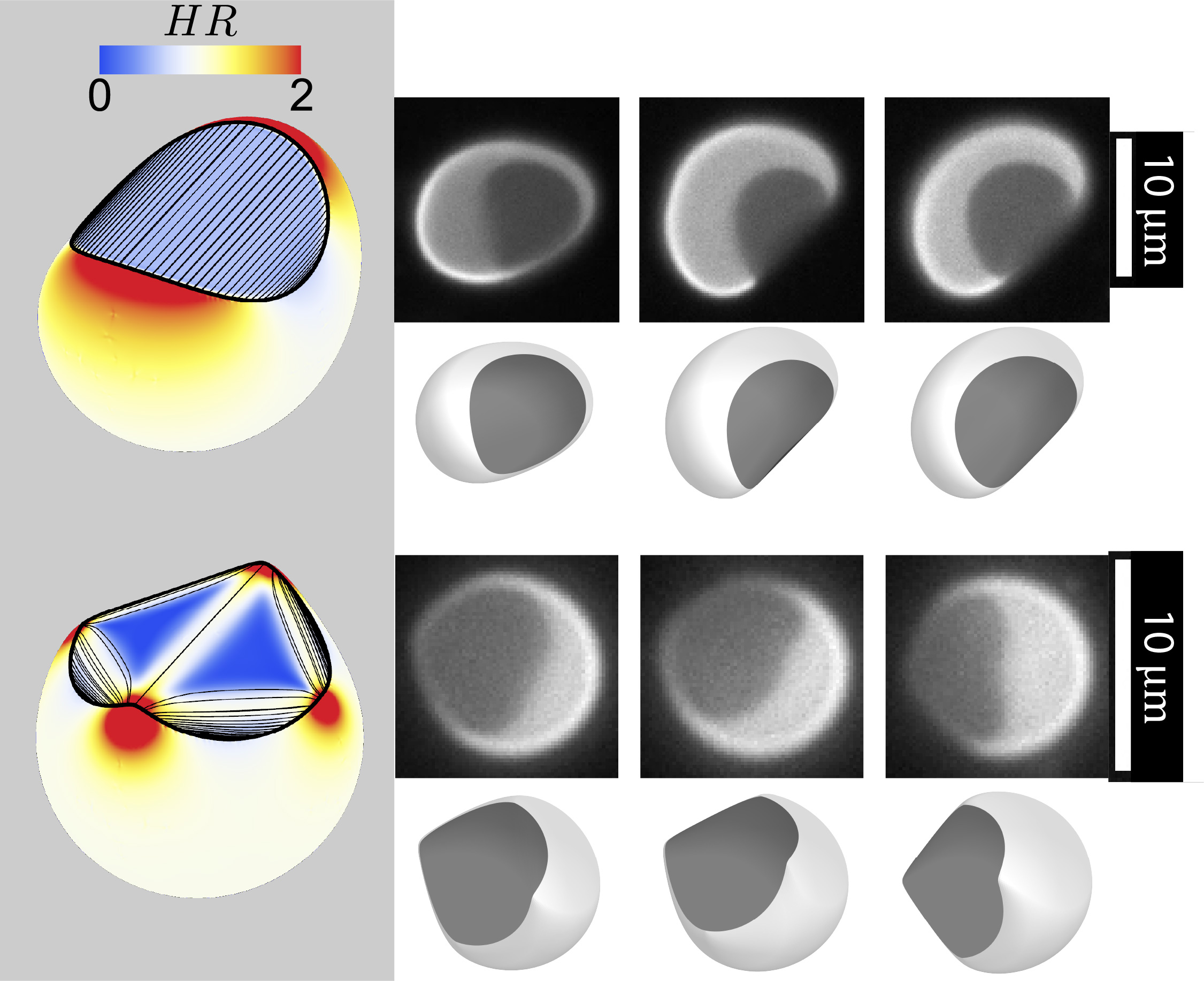}
\caption{\label{fig: experiment}\textbf{Experimental examples of roll and fold morphologies.} Confocal microscopy images of fluid-solid (DOPC-DPPC) composite GUVS are compared with multiple 2D projections of the Surface Evolver model for both the roll and fold phases. The simulated shapes are derived from model parameters ($\bar{t}^{-1}\approx 8.6\times 10^{3}$ consistent with the GUV size) using $\Phi=0.18, \bar{v}=0.95$ for the rolled case and $\Phi=0.18, \bar{v}=0.97$ for the folded case.}
\end{center}
\end{figure}

Taking a modest size for GUVs $R \approx 2~{\rm \mu m}$ and a generously large range for bending stiffness $B \approx 100 ~ k
_{\rm B} T$ shows that the bending tension scale is quite small $\gamma_{\rm bend} \approx {\rm \mu N / m}$, even well below the expected contributions of thermal fluctuations to tension~\cite{shi2014dynamics}.  This implies that even fluid-solid composite vesicles that are only very weakly inflated, provided they are at the GUV scale, may be well above the internal pressure and membrane tension scale needed to drive transitions to inhomogenously folded states, like polygonal-folds.  As a test of this prediction, we analyzed florescence microscope videos of DOPC-DPPC composite GUVs, prepared to include a single solid domain of roughly $\Phi \approx 0.16\pm0.02$ solid fraction.  Methods to prepare and image these vesicles are detailed in Appendix~\ref{app: experiment}, and are similar to what has been reported elsewhere~\cite{wan2024thermal}.  Notably, the florescent tracer dye is expelled from solid DPPC domain, and hence, only fluid DOPC-rich domain is visible in the imaging.  

Two extracted examples of these composite vesicles are shown in Figure~\ref{fig: experiment}, with multiple distinct frames of the movie showing somewhat different view directions of the vesicle [\href{https://drive.google.com/file/d/1YwoX7Uxa8jUAkyH5F1osLwgn9m4-i2KF/view?usp=drive_link}{see also supplementary video}].
Because fluorescence selectively labels the highly compliant fluid phase, the full 3D structure of the dark solid domain cannot be resolved directly; instead, its morphology must be inferred from the geometry of the fluid-solid boundary. By comparing all 2D image frames of the experimental videos, we find that these vesicles are well described as respective roll ($\Phi=0.18, \bar{v}=0.95$) and polgonal-fold ($\Phi=0.18, \bar{v}=0.97$) morphologies.  Notably, shape matching requires values of the inextensibility of the GUV size range (i.e. $\bar{t}^{-1} \approx 8.6\times10^{3}$), and the sharp, localized kinks at the domain boundaries evident from the microscopic of the polygonally-folded domain cannot be captured for the effectively extensible regime of LUVs.   This observation for sharply folded shapes in composite vesicles, and its strong match to isometrically folded morphologies of the composite fluid-solid elastic model, provide preliminary, yet direct, evidence of the impacts of 2D solid mechanics to drive strongly symmetry-breaking configurations.  A more systematic experimental study with controlled membrane tension, say via osmotically swollen vesicles or micromanipulated composite vesicles, is needed to confirm the predicted reversibility and tension threshold for transitions between smooth and folded states.  We note further that the predicted threshold to crumpled shapes, while larger than the threshold for folding, may also be in an experimentally accessible range of membrane tensions (and internal pressures), however, it is not possible to discern evidence of fine features like crumpling or wrinkling of the solid domains in this DOPC-DPPC system.

The 2D elastic properties of solid lipid bilayer phases (like DPPC domains) are not at present well characterized.  Nonetheless, we may assess a lower bound on 2D Young's modulus $Y$ from the characterized area expansion elasticity of membrane phases which ranges of order $10^3~ {\rm mN/m}$ for solid or gel DPPC phases~\cite{tierney2005elasticity,duncan2008comparing,drabik2020mechanical}, which for $\Phi \approx 0.1$ suggests strain tension scale of $\gamma_{\rm strain} \approx 0.1~{\rm N/m}$.  This tension scale is well above of what is typically expected for membrane lysis, $\sim 6-8$~mN/m~\cite{olbrich2000water,portet2010new,rawicz2008elasticity}, implying that composite membranes likely rupture before fully inflating into cap-like shapes.  That said, the threshold tension for wrinkling is found to be much smaller $\approx 10^{-2} \gamma_{\rm strain} $, which is conceivably in a range below lysis tension.  Again, it is not possible to observe evidence of fine-pattern features like wrinkles in the DOPC-DPPC vesicles shown in Figure~\ref{fig: experiment}.  Additional microscopy approaches from resolving sub-micron topographical features of solid domains are likely required to assess and quantify possibly wrinkled domains.

Throughout most of the article, we focused primarily on the case of uniform bending rigidity across both the fluid and solid domains ($\beta= 1$). While this assumption successfully isolated the role of dimensionless thickness and in-plane shear in driving morphological transitions, real lipid vesicles may exhibit a mechanical mismatch between phases. Experimental measurements of solid-phase bending rigidity remain technically challenging due to the suppression of thermal undulations in the solid state. However, recent atomic force microscopy indentations and molecular dynamics simulations predict that the highly ordered, extended lipid tails in the solid phase yield a bending stiffness roughly an order of magnitude larger than that of the fluid phase ($\beta \sim 10$)~\cite{et2017mechanical,thakkar2011verifying,diggins2015curvature,ridolfi2021stiffness}. To address this physical reality, we extend our analysis in Appendix~\ref{app: bend} to map the morphological phase diagram as a function of inflation $\bar{v}$ and solid fraction $\Phi$ for various rigidity ratios $\beta > 1$. We find that introducing even a modest excess bending stiffness to the solid domain significantly flattens it, as demonstrated in Figure \ref{fig: zero_tension-pressure}. Consequently, composite vesicles with $\beta > 1$ strongly prefer to embed the solid domain on the flat poles of an oblate shape branch, fundamentally suppressing the prolate configurations favored when $\beta = 1$.

Last, solid domains may take on many complex, non-circular 2D shapes in vesicles. Indeed, it is well-known that geometric frustration of 2D solid forming on surface with positive Guassian curvature tends to favor extended and non-convex shapes, such as ribbon-like strips or branched structures in order to avoid the elastic penalties of geometric frustration~\cite{starostin2007shape,witten2007stress}.  We have previously shown that under sufficient processing conditions during solidification, solid DPPC domains may grow into an elaborate spectrum of non-convex shapes, from hexagonally faceted domains to 6-fold symmetry flowers efficiently conform to spherical shapes of highly inflated vesicles~\cite{wan2024flower,wan2024thermal,wan2026sculpting}.  This raises an open line of question about how shape transitions studied here for a fixed class of circular 2D solid domains is altered by a non-axisymmetry of the solid domain itself.  Based on prior simulations that show non-convex solid domains conform to nearly fully inflated (i.e. $\bar{v} \approx 1$) shapes at much lower elastic penalty than is possible for convex solid domains of the same area fraction, we expect that transitions to more complex 3D elastic patterns (e.g. folded, crumpled or wrinkled) may be pushed to an even higher tension scale, if they are not eliminated altogether.  Furthermore, it is likely for non-convex solid domain shapes like flowers that the selected elastic patterns themselves will couple strongly to the anisotropy the domain shape, e.g. folding induced at the stems of petal-like features.  Effects of this type have been studied, and engineered, in the context of so-called capillary origami wrapping of non-circular sheets into low symmetry, e.g. polyhedral, liquid droplets, but their broad implications for collective shape and mechanical control of composite vesicles remains to be systematically unexplored.

\section*{Acknowledgements}
We thank B. Davidovitch for useful discussions about this work.  This work was supported by the U.S. Department of Energy, Office of Science, Basic Energy Sciences, under award DE-SC0017870.  Surface Evolver simulations were performed on the Unity Cluster at the Massachusetts Green High Performance Computing Center

\bibliographystyle{unsrt} 
\bibliography{reference}

\appendix

\section{Surface Evolver minimization}
\label{app: SE}

The computations and simulations were performed using the computational software Surface Evolver (SE). The mesh was initialized with $\approx 44,000$ vertices. The fluid area was fixed as a constraint, and the area of the unstrained, flat solid domain was determined by a target solid fraction, $\Phi_\text{target}$, and subsequently optimized via energy minimization. The enclosed volume was varied for a given target total area $A_\text{target}$ to establish the target reduced volume $\bar{v}_\text{target}$.

The membrane tension and pressure difference values are extracted from the Lagrange multipliers conjugate to fluid area constraint and enclosed volume constraint respectively.   Note that extraction of pressure and tension from Surface Evolver becomes very sensitive to small changes in mesh geometry, notably at large inflation.  In particular, we find that extracted values of tension are numerically unstable following the use of ``hessian'', or ``hessian\_seek'', both of which are used to search for global minima.  Therefore, after full equilibration and before extracting the final tension, we apply several steps ``conjugate gradient'' descent until tension values are numerical stable.

The bending energy was computed using the built-in \texttt{star\_perp\_sq\_mean\_curvature} function applied to the global vertices, while the solid strain energy was imposed using the \texttt{linear\_elastic} function applied to the solid facets. In the modest to thin-sheet regimes ($\bar{t}^{-1} \gtrsim 10^2$), the solid fraction remains nearly constant throughout most of the reduced volume range. However, in the thick-plate regime ($\bar{t}^{-1} \lesssim 10^2$) and under extreme inflation ($\bar{v} \approx 1$), the solid domain area can dilate, incurring a strain energy penalty. The total elastic energy was minimized using gradient descent and Hessian methods until the system was fully equilibrated for a given set of physical constraints: $\Phi_\text{target}$, $\bar{v}$, and $\bar{t}^{-1}$. Auxiliary commands, such as \texttt{vertex averaging} and \texttt{jiggle}, along with manual tuning of the initial configuration, were employed as needed~\cite{brakke1992surface,brakke2013surface}. The preparation of the initial mesh and the energy minimization protocol were described in detail in~\cite{jeon2024shape}, and briefly summarized below.

\begin{itemize}
    \item[1)] Prepare an isoperimetric limit configuration by projecting down a portion of a spherical mesh, and identify the flat portion as the initially unstrained solid domain, and the rest as fluid membrane.
    
    \item[2)] Set the target attributes such as fluid area, total volume, bending modulus, and strain modulus to match $\Phi_\text{target}$, $\bar{v}_\text{target}$, and $\bar{t}^{-1}$. To impose different bending moduli for the solid and fluid domains ($\beta > 1$), we apply a baseline global bending modulus $B_f$ to all vertices, and then add an excess bending modulus $B_\text{s} - B_\text{f}$ specifically to the solid vertices. Applying a global bending background, rather than entirely separate and disjoint moduli for the fluid and solid, is necessary to ensure geometric smoothness across the boundary.
    
    \item[3)] Relax the total elastic energy using \texttt{conjugate\_grad} and \texttt{hessian\_seek} until the step size falls below a critical threshold ($10^{-9}$).
\end{itemize}

\section{Shape Classification Algorithm}
\label{app: classification}

To systematically categorize the morphological transitions of the simulated fluid-solid composite vesicles, we implemented an automated shape classification algorithm. The algorithm evaluates the discrete mesh at each structure to assign one of six morphological states: roll, edge-fold, polygonal-fold, crumple, wrinkle, or Cap. 

Rather than relying on visual inspection, the classification utilizes a hierarchical decision tree based on geometric profiles: The algorithm first tests if the structure is smoothly developable (low angle defect, moderate $H_\text{min}$). Smoothly developable shapes are branched into rolls, edge-folds, or polygonal-folds based on their curvature contrast and boundary modes. If the structure exhibits sharp, localized deformation but low global strain, it is categorized as a Crumple. Finally, stretched, non-developable states are assessed for Wrinkles via boundary peak counting; if no distinct boundary waves are present, the structure defaults to a nominally spherical cap. Full implementation details, including the exact quantitative thresholds, coordinate alignment, and dynamic detrending logic, are provided in the accompanying Mathematica notebook.


\section{Wrinkle analysis: Far-from-threshold elastic plate theory}
\label{app: wrinkle}

In the highly bendable (thin-sheet) regime, the morphological features of the wrinkled state can be analytically described using far-from-threshold (FFT) asymptotic expansions of the Föppl--von Kármán equations. Detailed theoretical derivations of this framework for a circular sheet on a deformable spherical drop can be found in previous literature~\cite{king2012elastic, paulsen2016curvature}. Here, we adapt this theory to our system, utilizing the dimensionless reduced tension $\tilde{\gamma} = \gamma / Y \Phi$, which is inversely proportional to the confinement parameter $\alpha$ used in those previous works (where $\tilde{\gamma} \approx 1/2\alpha$).

\subsection{Wrinkle extent}

In the FFT limit, the wrinkles act to completely relax the compressive hoop stress, rendering the stress field compression-free ($\sigma_{\theta\theta} = 0$) in the wrinkled annulus ($W - L < r < W$, where $L$ is the radial distance from the center of the domain to the wrinkled zone). Imposing this condition reduces the in-plane radial force balance, $\partial_r(r\sigma_{rr}) - \sigma_{\theta\theta} = 0$, to simply $\partial_r(r\sigma_{rr}) = 0$. Applying the boundary condition at the perimeter ($\sigma_{rr}(W) = \gamma$) yields a $1/r$ decay for the radial stress: $\sigma_{rr}(r) = \gamma W / r$.

This wrinkled outer annulus acts as a boundary condition for the unwrinkled inner core ($r < W - L$). Due to the nonlinear coupling of the radial stress in the Föppl--von Kármán equations, the effective reduced tension experienced by this inner core scales as $\tilde{\gamma}_\text{eff}\sim [\sigma_{rr}(W-L)/\gamma]^3[W/(W-L)]^2 \tilde\gamma$. Substituting $\sigma_{rr}(W-L) = \gamma W / (W-L)$ reveals that the effective reduced tension grows rapidly as $[W/(W-L)]^5\tilde{\gamma}$. 

The boundary of the wrinkled zone is determined by the dominant balance condition: this effective reduced tension must exactly match the critical threshold ($\tilde{\gamma}_0$) for a stable, unwrinkled state. Equating $\tilde{\gamma}[W/(W-L)]^5 = \tilde{\gamma}_0$ naturally yields the power-law scaling $(W-L)/W \approx (\tilde{\gamma}/\tilde{\gamma}_0)^{1/5}$. Therefore, the scaling behavior for the relative extent of the wrinkle zone is given by:
\begin{equation}
    1-\frac{L}{W} \approx \left(\frac{\tilde{\gamma}}{\tilde{\gamma}_0}\right)^{1/5}
\end{equation}
where $\tilde{\gamma}_0 \approx 0.4$ is the critical reduced tension at which the wrinkle zone completely vanishes (i.e., the state becomes fully axisymmetric and $L=0$).

\subsection{Number of wrinkles}

The local wavelength of the wrinkles, $\lambda$, is determined by a balance between the bending modulus $B$, which resists curving, and an effective stiffness $K_\text{eff}(r)$ that resists out-of-plane deformation~\cite{paulsen2016curvature}:
\begin{equation}
    \lambda=2\pi \left(\frac{B}{K_\text{eff}}\right)^{1/4}
\end{equation}
The effective stiffness is generally composed of three terms: a substrate stiffness ($K_\text{sub}$), a tension-induced stiffness ($K_\text{tens}$), and a curvature-induced stiffness ($K_\text{curv}$). In our system, the gravitational substrate energy is negligible. Furthermore, deep in the FFT regime (where the strain modulus dominates, $YW^2/\gamma R^2 \gg 1$, or equivalently $\tilde{\gamma} \ll 1$), the curvature-induced stiffness overwhelmingly dominates the tension-induced stiffness ($K_\text{curv} \gg K_\text{tens}$) over the vast majority of the wrinkled zone. 

Therefore, we can approximate $K_\text{eff} \approx K_\text{curv} \approx Y (h_0'')^2$, where $h_0$ is the axisymmetric normal displacement. For a small domain on a large sphere ($W \ll R$), the radial curvature scales as $h_0'' \approx 2r/RW$. Thus, the effective stiffness is given by:
\begin{equation}
    K_\text{eff}(r) \approx Y\left(\frac{2r}{RW}\right)^2
\end{equation}
The local number of wrinkles $m(r)$ at a radial distance $r$ is geometrically related to the wavelength by $m(r) = 2\pi r / \lambda$. Substituting $K_\text{eff}(r)$ into the wavelength expression, we find that the local number of wrinkles scales as:
\begin{equation}
    m(r)=\frac{2\pi r}{\lambda}\approx 2^{3/2}\sqrt{\Phi}\; \bar{t}^{-1/2}\left(\frac{r}{W}\right)^{3/2}
\end{equation}
This leads to the spatial scaling rule for the wrinkle density, normalized by the solid fraction:
\begin{equation}\label{eq: wrknumbers}
    \frac{m}{\sqrt{\Phi}} \sim \bar{t}^{-1/2}\left(\frac{r}{W}\right)^{3/2}
\end{equation}
This $3/2$ power-law dependence on the radial coordinate indicates that the wrinkles become denser and more sharply defined toward the outer boundary of the solid domain, gradually cascading into smoother waves as they approach the unwrinkled core.

\subsection{Mesh Resolution Limits for Hierarchical Wrinkling}
\label{app: mesh_resolution}

To ensure the numerical validity of the simulated wrinkling morphologies, we must establish the theoretical upper bound of wrinkle modes supported by the discretized mesh. As observed in Figure~\ref{fig: non-isometry}(B), the number of wrinkles saturates at extreme inverse thicknesses ($\bar{t}^{-1}\sim 10^4$). This saturation is a geometric artifact corresponding to the spatial resolution limit of the finite element model.

Our vesicle models are constructed from a uniform spherical triangulation of $\approx 44{,}000$ vertices. Based on this resolution, the number of vertices residing on the solid domain boundary scales as $\approx 700 \sqrt{\Phi}$. Requiring a minimum of two vertices to geometrically resolve a single physical wave, the maximum normalized number of wrinkles supported strictly at the boundary ($r = W$) is $m(W)/\sqrt{\Phi} \approx 350$. 

However, the hierarchical nature of the wrinkles dictates that this resolution degrades rapidly toward the domain interior. Following Equation~\ref{eq: wrknumbers}, the number of physical wrinkles $m(r)$ follows a $3/2$ power law decay from the boundary inward. Therefore, the representative numerical capacity of the mesh over the primary wrinkling domain ($r \in [0.5W, W]$) is best captured by the 1D radial average, $\langle m \rangle/\sqrt{\Phi}$:
\begin{equation}
    \frac{\langle m \rangle}{\sqrt{\Phi}} \approx \frac{1}{0.5W} \int_{0.5W}^{W}\d r\  \frac{m(W)}{\sqrt{\Phi}} \left( \frac{r}{W} \right)^{3/2} \approx 230
\end{equation}
This theoretical radial capacity closely aligns with the empirical saturation point observed in our simulations ($\langle m \rangle/\sqrt{\Phi} \sim 200$), confirming that the truncation of the scaling trend at $\bar{t}^{-1}\sim 10^4$ is entirely due to the discrete spatial resolution of the mesh, rather than a physical breakdown of the far-from-threshold theory.

Consequently, simulations exhibiting an average normalized wrinkle count approaching or exceeding $\sim 230$ are constrained by discrete mesh resolution, which may artificially stiffen the sheet and suppress higher-order wrinkling modes.


\section{Role of solid bending stiffness}\label{app: bend}

In the preceding sections, we assumed identical bending moduli for both the fluid and solid phases ($\beta=1$). This simplification allowed us to isolate the impact of strain elasticity---a feature entirely absent in homogeneous fluid membranes. The morphological role of heterogeneous bending moduli has already been systematically explored in the context of multiphase fluid vesicles~\cite{julicher1996shape, baumgart2003imaging, gutlederer2009polymorphism}, where the structural impact is relatively straightforward compared to the geometric frustration induced by solid shear elasticity. Nevertheless, understanding the effect of relative bending stiffness is essential for drawing comparisons with experimental vesicle systems. 

Although precise experimental measurements of the bending modulus of solid-phase membranes remain technically challenging due to the suppression of thermal undulations in the gel state, recent atomic force microscopy (AFM) indentations~\cite{et2017mechanical, ridolfi2021stiffness} and molecular dynamics simulations~\cite{diggins2015curvature, thakkar2011verifying} have successfully quantified this mechanical mismatch. These studies consistently indicate that the highly ordered lipid tails in solid domains yield a bending stiffness roughly an order of magnitude larger than that of the surrounding fluid membrane. Specifically, AFM measurements report the Young's modulus jumping from $\sim 13$~MPa in fluid DOPC to $\sim 116$~MPa in gel-phase DPPC~\cite{et2017mechanical}, while molecular dynamics simulations independently predict a $10$- to $15$-fold increase in the bending modulus for the solid phase~\cite{diggins2015curvature, thakkar2011verifying}. 

\begin{figure*}[t!]
\begin{center}
    \includegraphics[width=\textwidth]{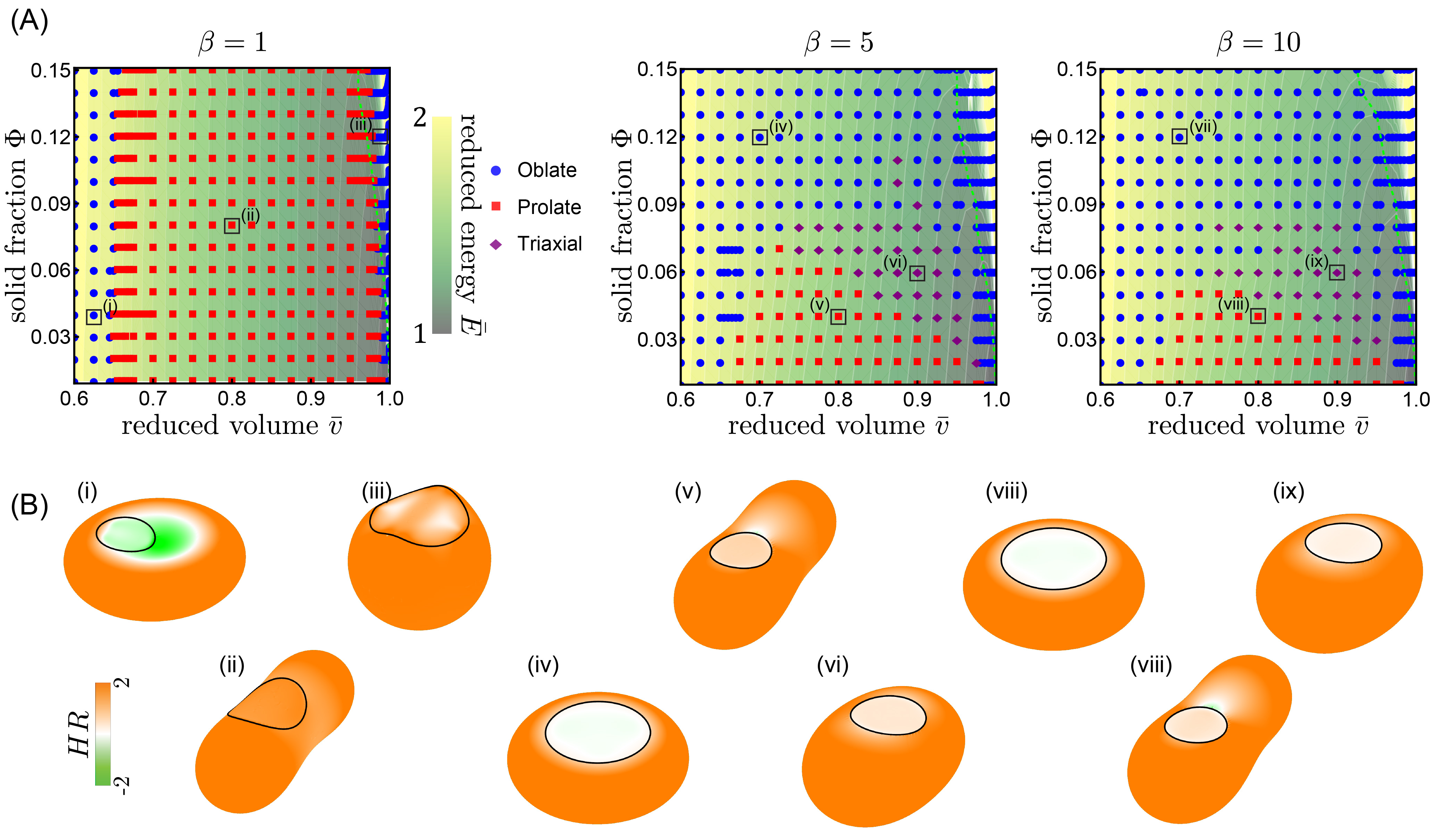}
\caption{\label{fig: beta}\textbf{Phase diagram of various solid-to-fluid bending stiffness ratio.} (A) Phase diagram of shape equilibria for various bending stiffness ratios $\beta$ with a fixed inverse thickness of $\bar{t}^{-1} \approx 2.7\times 10^{3}$ and with representative morphological configurations shown in (B), shaded according to curvature.}
\end{center}
\end{figure*}

To address this physical reality, we investigate shape equilibria by varying the relative bending stiffness ratio $\beta$ in the GUV thin-sheet regime ($\bar{t}^{-1} \approx 8.6\times 10^{3}$).

Figure~\ref{fig: beta} presents the morphological phase diagram and representative structures for $\beta=1, 5,$ and $10$, keeping $\bar{t}^{-1} \approx 8.6\times 10^{3}$. For $\beta=1$---where the solid bending rigidity matches that of the fluid and in-plane strain effects are the primary driver of geometric frustration---the vesicles strongly prefer prolate shapes within the reduced volume range $0.655 \lesssim \bar{v} \lesssim 1$. Conversely, for $\beta > 1$, the solid domain rapidly flattens, exhibiting a strong geometric incompatibility with the high-curvature waist of prolate shapes. This mechanical mismatch shifts the global shape preference heavily toward the oblate or triaxial branches, where the stiffer solid domain can reside on relatively flatter regions of the vesicle. As a result, for $\beta \gtrsim 10$, the morphological phase diagram closely approaches the rigid planar domain limit previously established in~\cite{jeon2024shape}.

\section{Experiment details}
\label{app: experiment}

Giant Unilamellar Vesicles (GUVs) were produced from a mixture of 1,2-dioleoyl-sn-glycero-3-phosphocholine (DOPC) and 1,2-dipalmitoyl-sn-glycero-3-phosphocholine (DPPC) lipid solutions with fluorescent tracer lipid, 1,2-dioleoyl-sn-glycero-3-phosphoethanolamine-N-(lissamine rhodamine B sulfonyl) (ammonium salt) (Rh-DOPE) using the electroformation method as previously described~\cite{xin2021switchable,wan2024flower}. The DOPC, DPPC, and fluorescent tracer lipids Rh-DOPE were purchased from Avanti Polar Lipids (Alabaster, AL). 

Vesicles composed of DPPC/DOPC at a molar ratio of 25/75 plus 0.1 mol\% Rh-DOPE were electroformed on platinum wires. Before electroformation, a 100 mM sucrose solution was preheated and used as the electroforming solution. To ensure vesicles formed were in the one-phase region of the phase diagram, the electroforming temperature was maintained in the range 60--65~$^\circ$C for 1 hour, with an applied frequency of 10 Hz and voltage of 2 V. After electroformation, the stock vesicle suspension was harvested and allowed to cool to room temperature for later use. Vesicles with single solid domains were formed by dilution of the harvested suspension in 100 mM sucrose solution, placed into a $24~\text{mm} \times 60~\text{mm}$ closed chamber made from two coverslips and parafilm spacers, reheated to 55~$^\circ$C for 15 minutes, and subsequently cooled at a controlled rate of 0.3~$^\circ$C\,min$^{-1}$ from 55~$^\circ$C to room temperature. 

While the numbers of domains per vesicle was variable between 1--2, depending on vesicle size, vesicles having single solid domains were imaged using a Nikon Eclipse TE 300 inverted epifluorescence microscope equipped with a 40$\times$ long working distance air fluorescence objective. Images were recorded with a pco.panda 4.2 sCMOS monochrome camera and analyzed using Nikon NIS Elements imaging software.

\end{document}